\documentclass[12pt]{article}
\usepackage[latin1]{inputenc}

\usepackage{amsmath}
\usepackage{color}
\usepackage{amsfonts}
\usepackage{amssymb}
\usepackage[mathscr]{euscript}
\usepackage{graphicx}
\usepackage{geometry}
\usepackage{amssymb,epsfig}
\usepackage{hyperref}
\usepackage{comment}
\usepackage{tabularx}
\usepackage{bm}
\usepackage{euscript}
\usepackage{graphicx}
\usepackage[dvipsnames]{xcolor}
\usepackage{amsfonts}
\usepackage{exscale}
\usepackage{amsbsy}
\usepackage{subcaption}
\usepackage{textcomp}
\usepackage{hyperref}
\hypersetup{colorlinks,allcolors=black}
\usepackage{slashed}
\usepackage{authblk}
\usepackage{tabularx}
\usepackage{euscript}
\usepackage{graphicx}
\usepackage{color}
\usepackage{exscale}
\usepackage{amsbsy}
\usepackage{textcomp}
\usepackage{doi}
\usepackage{tensor}
\usepackage{wrapfig}
\usepackage[font=footnotesize,labelfont=bf]{caption}
\usepackage{makecell}
\usepackage[mathscr]{euscript}

\def\ba#1\ea{\begin{align}#1\end{align}}
\def\bg#1\eg{\begin{gather}#1\end{gather}}
\def\bm#1\em{\begin{multline}#1\end{multline}}
\def\bmd#1\emd{\begin{multlined}#1\end{multlined}}

\newcommand{\be}{\begin{equation}}
	\newcommand{\ee}{\end{equation}}
\newcommand{\bea}{\begin{eqnarray}}
	\newcommand{\eea}{\end{eqnarray}}

\newcommand{\matleft}{\left(\begin{array}}
	\newcommand{\matright}{\end{array}\right)}
\newcommand{\Tr}{\operatorname{Tr}}

\newcommand{\sgn}{\operatorname{sgn}}
\newcommand{\dd}{\mathrm{d}}

\usepackage[numbers,sort&compress]{natbib}
\def\simge{
	\mathrel{\rlap{\raise 0.511ex 
			\hbox{$>$}}{\lower 0.511ex \hbox{$\sim$}}}}

\def\simle{
	\mathrel{\rlap{\raise 0.511ex 
			\hbox{$<$}}{\lower 0.511ex \hbox{$\sim$}}}}

\makeatletter
\renewcommand\section{\@startsection {section}{1}{\z@}%
	{-3.5ex \@plus -1ex \@minus -.2ex}
	{2.3ex \@plus.2ex}%
	{\normalfont\large\bfseries}}
\renewcommand\subsection{\@startsection{subsection}{2}{\z@}%
	{-3.25ex\@plus -1ex \@minus -.2ex}%
	{1.5ex \@plus .2ex}%
	{\normalfont\bfseries}}
\renewcommand\subsubsection{\@startsection{subsubsection}{3}{\z@}%
	{-3.25ex\@plus -1ex \@minus -.2ex}%
	{1.5ex \@plus .2ex}%
	{\normalfont\itshape}}
\makeatother

\def\pplogo{\vbox{\kern-\headheight\kern -29pt
		\halign{##&##\hfil\cr&{\ppnumber}\cr\rule{0pt}{2.5ex}&\ppdate\cr}}}
\makeatletter
\def\ps@firstpage{\ps@empty \def\@oddhead{\hss\pplogo}%
	\let\@evenhead\@oddhead 
}
\hypersetup{
	unicode=false,          
	pdftoolbar=true,        
	pdfmenubar=true,        
	pdffitwindow=false,     
	pdfstartview={FitH},    
	pdftitle={CI boundary criticality},    
	pdfauthor={},     
	pdfsubject={Subject},   
	pdfcreator={},   
	pdfproducer={}, 
	pdfkeywords={keyword1} {key2} {key3}, 
	pdfnewwindow=true,      
	colorlinks=true,       
	linkcolor=OliveGreen, 
	citecolor=NavyBlue,        
	filecolor=magneta,      
	urlcolor=cyan           
}

\numberwithin{equation}{section}

\newcommand*\samethanks[1][\value{footnote}]{\footnotemark}

\newcommand\beal{\begin{equation}\begin{aligned}}
		\newcommand\eeal{\end{aligned}\end{equation}}

\begin{document}
	
	
	\setcounter{page}0
	\def\ppnumber{\vbox{\baselineskip14pt
	}}
	
	\def\ppdate{
	} 
	\date{\today}

	\title{\Large\bf Chern insulator boundary criticality}
	\author{Benjamin Moy and Eduardo Fradkin}
	\affil{\it\small Department of Physics and Anthony J. Leggett Institute for Condensed Matter Theory,\\\it\small University of Illinois Urbana-Champaign, Urbana, Illinois 61801, USA}
	\maketitle\thispagestyle{firstpage}
	\begin{abstract}

  We investigate signatures of chirality at Chern insulator transitions in the presence of a boundary. The transition between a trivial insulator and a Chern insulator with Chern number $C=1$ is described by a massless Dirac fermion whose parity anomaly gives a critical Hall conductivity $\sigma_{xy}=\frac{1}{2}\frac{e^2}{h}$. Using a Dirac mass domain wall construction, we show that the chiral edge mode delocalizes into the bulk at criticality, but the boundary fermion correlation function retains a chiral structure and acquires the scaling dimension of the bulk fermion. We compute current correlation functions and demonstrate that the anomaly of the bulk Hall response is matched by delocalized chiral modes near the boundary. Using only the residual conformal symmetry of a (2+1)d conformal field theory (CFT) in a half-space, we identify parity-odd terms in current and energy-momentum tensor correlation functions that encode these modes and determine their electromagnetic and gravitational anomaly coefficients. Our analysis therefore applies to general time-reversal breaking (2+1)d CFTs, beyond the free Dirac transition. In particular, the analysis of the gravitational anomaly is also applicable to the free Majorana CFT governing the transition between a trivial superconductor and a topological superconductor. We also extend our results to more general Chern number changing transitions and to the transition between a (3+1)d topological insulator and a trivial insulator.

	\end{abstract}
	
	\pagebreak
	{
		\hypersetup{linkcolor=black}
		\tableofcontents
	}
	\pagebreak

\section{Introduction}

Topological phases of matter are often most sharply characterized by their boundaries. In quantum Hall systems, the universal Hall response of the bulk is accompanied by chiral edge modes, whose correlation functions encode universal information about the topological phase~\cite{Wen-1990h,Wen-1990f,Wen-1995}. This bulk-edge correspondence is usually formulated for gapped phases, where the low-energy boundary degrees of freedom are sharply separated from the bulk. At a continuous transition between two topological phases, however, the bulk gap closes. The edge can then couple to the critical bulk modes, and it is no longer obvious what becomes of the usual chiral boundary theory.

Several questions naturally arise. When the Chern number changes at a continuous quantum Hall transition, do the chiral edge modes remain localized near the boundary, or do they delocalize into the bulk? If the bulk critical point is not time-reversal invariant, as in quantum Hall transitions (or quantum anomalous Hall transitions), what observable signatures of chirality remain at the edge? Finally, how does anomaly inflow work when the degrees of freedom that match the anomaly are not sharply localized at the boundary? These questions are closely related to recent work on gapless symmetry-protected topological phases (SPTs)~\cite{Keselman2015,Scaffidi2017,Jiang2018,Parker2018,Verresen2018,Thorngren2021,Wen2023,Yu2026}, where symmetry and anomaly matching can constrain boundary phenomena even when the bulk is gapless. Nevertheless, a general understanding of topology and edge modes in gapless systems remains incomplete. There is also a long history of boundary criticality in symmetry-breaking transitions~\cite{Mills1971,Binder1972,Binder1974,Bray1977a,Bray1977b,Ohno1983,Ohno1984,McAvity1995,Diehl1997}, which has recently been revived because of a new extraordinary log boundary universality class~\cite{Metlitski2022} and has been extended to a variety of contexts~\cite{Sun2022,Zou2022,Krishnan2023,Lee2023,Cuomo2024,Sun2025,Cui2026}. By contrast, much less is known about critical points with chiral boundary states, which naturally arise in the quantum Hall setting, and will be our focus here.

In this work, we address these questions in the simplest setting where they can be studied explicitly: the transition between a trivial insulator and a Chern insulator with Chern number $C=1$. Chern insulators provide a lattice realization of the quantum Hall effect with no external magnetic field~\cite{Haldane1988}. The transition we study is described at long distances by a single massless Dirac fermion in 2+1 dimensions. Because of the parity anomaly~\cite{Redlich1984a,Redlich1984b}, a proper regularization of the critical Dirac fermion carries a half-integer Hall response $\sigma_{xy}=e^2/2h$. Equivalently, the critical point has an
effective Chern number $C=1/2$. Thus, regardless of whether there is a localized chiral edge mode at the critical point, since the bulk critical point is not time-reversal invariant, there should be physical signatures of chirality at the edge of the system.

To study the boundary physics of the transition, we employ a fermion mass domain wall construction~\cite{Jackiw1976,Callan1985,Kaplan1992}. The material occupies the half-space $y<0$ while the region $y>0$ is treated as vacuum (or a trivial insulator). At the critical point, the Dirac mass vanishes for $y<0$ and is nonzero for $y>0$. In the limit where the absolute value of the mass in the region $y>0$ is taken to infinity, the problem reduces to a massless Dirac fermion in a half-space, $y<0$, with a chiral boundary condition. This construction is closely related to the usual domain wall description of chiral edge modes in the gapped Chern insulator phase.

Our first main result is the fermion two-point function at the critical point. At the boundary of the system, this correlation function reduces to
\begin{equation}
\label{eq: bdry propagator intro}
    \left\langle \mathcal{T}[\psi(t,x,y=0)\, \bar{\psi}(0)]\right\rangle= \left(\gamma^t-\gamma^x\right)\frac{1}{4\pi}\frac{1}{\sqrt{t^2-x^2}}\frac{1}{x-t},
\end{equation}
where $\mathcal{T}$ denotes time-ordering, $\gamma^\mu$ are Dirac matrices, and we work in units in which the Fermi velocity is $v_F=1$. This correlation function has the same chiral Dirac matrix structure as the edge propagator of a gapped Chern insulator, but the critical bulk dresses the chiral fermion with the factor $\left(t^2-x^2\right)^{-1/2}$, modifying the boundary chiral fermion observables and introducing an anomalous dimension. The boundary fermion retains the chiral nature of the edge mode in the gapped phase while acquiring the scaling dimension of the bulk Dirac fermion. Thus, although the chiral edge mode delocalizes into the bulk at criticality~\cite{Jansen1992}, observable signatures of chirality remain at the edge. This unusual boundary correlation function may be regarded as a distinct chiral boundary universality class.

In the Chern insulator phase, the boundary supports a conventional chiral edge mode that is exponentially localized near the interface. To determine how this mode delocalizes and evolves into Eq.~\eqref{eq: bdry propagator intro} at the transition, we study fermion and current correlation functions as the critical point is approached from the Chern insulator phase. As the gap closes, the length scale governing the decay of the edge mode into the bulk diverges. At criticality, the exponentially decaying profile is replaced by a power law. While this chiral mode is no longer sharply distinct from bulk modes, its existence is due to the boundary.

Another central result of this work is the explicit demonstration of anomaly inflow at the critical point. In the Chern insulator phase, the anomaly of the bulk Chern-Simons response is canceled by a localized chiral edge mode~\cite{Callan1985,Golterman1993}. In contrast, at the critical point, we explicitly show that the delocalized chiral mode exactly matches the anomaly of the half-integer Hall conductivity. The anomaly and the delocalized chiral mode are encoded in parity-odd structures in the current and energy-momentum tensor correlation functions that, to our knowledge, have not previously been systematically studied. Importantly, the possibility of these terms can be deduced directly from the residual conformal symmetry of a conformal field theory (CFT) in the presence of a boundary. Thus, when the bulk is gapless, the anomaly need not be matched solely by boundary degrees of freedom; it may instead be matched in part (or entirely) by bulk critical modes near the system edge. Beyond the example of the Chern insulator transition, we use these symmetry-allowed structures to identify the delocalized chiral mode and calculate its electromagnetic and gravitational anomaly coefficients for a general (2+1)d CFT in a half-space. As we comment later, our analysis of the gravitational anomaly for the Chern insulator transition may be easily adapted to the free Majorana CFT that governs the transition between a trivial superconductor with chiral central charge $c_-=0$ and a $p_x+i p_y$ superconductor with chiral central charge $c_-=1/2$.

Finally, we generalize our results in two directions. First, we consider more general Chern insulator transitions between phases with Chern numbers $C=k$ and $C=k+\Delta k$. As discussed previously in Ref.~\cite{Verresen2020}, such transitions may host both delocalized chiral modes and chiral modes localized at the edge. Within our framework, we use gauge invariance and anomaly matching to show that the allowed combinations are constrained by the effective Chern number of the bulk critical point, giving a critical version of the bulk/edge correspondence. Second, we use our framework to study boundary physics at a transition between a (3+1)d time-reversal invariant topological insulator~\cite{Fu2007,Moore2007,Roy2009,Qi-2011} and a trivial insulator. At this transition, the surface fermion similarly acquires the scaling dimension of the bulk critical fermion, but the current correlation function remains time-reversal invariant, as expected for the topological insulator critical point.

We proceed as follows. In Section~\ref{sec: physical setup}, we introduce the lattice Chern insulator model, review the parity anomaly, and explain the mass domain wall construction used to model the boundary physics we wish to study. In Section~\ref{sec: two-point function}, we compute the fermion two-point function and examine its behavior near the edge of the system. In Section~\ref{sec: currents and anomalies}, we compute current correlation functions and use them to study anomaly inflow at the critical point. In Section~\ref{sec: thermal response}, we calculate correlation functions of the energy-momentum tensor and analyze anomaly inflow for gravitational response. In Section~\ref{sec: generalized chern transitions}, we generalize our results to other Chern insulator transitions. Finally, in Section~\ref{sec: TI boundary criticality}, we apply the same methods to analyze boundary physics at a transition between a (3+1)d topological insulator and a trivial insulator. We conclude in Section~\ref{sec: discussion} with a discussion of our results and possible future directions. 

\section{Physical setup: Dirac mass domain wall}

\label{sec: physical setup}

\begin{figure}

\centering\includegraphics[width=0.5\textwidth]{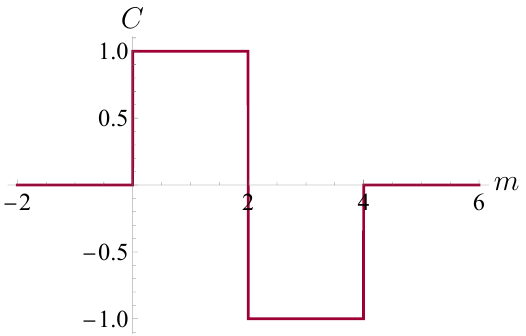}
	\caption{A plot of the Chern number $C$ for the lattice model Hamiltonian in Eq.~\eqref{eq:QHZ} as the 
    parameter $m$ is varied. In a given gapped phase, the Hall conductivity is $\sigma_{xy}=C\, e^2/h$. 
    For most of this work, we focus on the transition at $m=0$ between a Chern insulator with $C=1$ 
    and a trivial insulator.
    }
	\label{fig: chern number plot}
\end{figure}

We begin by describing the physical context of the quantum phase transition we wish to study. We consider a two-band tight-binding model on a square lattice with the Hamiltonian~\cite{Haldane1988,Qi2006,Wilson1974},\footnote{Strictly speaking, this is the Hamiltonian of \eqref{eq:QHZ} 
of the Qi-Hughes-Zhang model
    in two dimensions \cite{Qi2006} which has the pattern of Chern numbers shown in Eq.~\eqref{eq:masses}. 
    In the case of the Haldane model~\cite{Haldane1988} there is only one transition from Chern number $C=1$ 
    to a trivial insulator with $C=0$. The same holds for a square lattice with $\pi$ flux per plaquette. 
    Unlike Eq.~\eqref{eq:QHZ}, in these other cases there is only one heavy fermion doubler. 
    The more complex pattern of the Chern number shown in Eq.~\eqref{eq:masses} 
    for the model of Eq.~\eqref{eq:QHZ} is due to the contributions of additional heavy fermions at the corners 
    of the Brillouin zone becoming light at special values of the mass parameter $m$.}
\begin{align}
\label{eq:QHZ}
    \begin{split}
        H
        =\sum_{\mathbf{k}}c^\dagger(\mathbf{k}) \left[-\sin k_x \,\sigma^3+\sin k_y \, \sigma^2+\left(m-2+\cos k_x+\cos k_y\right)\sigma^1\right]c(\mathbf{k}).
    \end{split}
\end{align}
where $c^\dagger(\mathbf{k})$ creates an electron with (crystal) momentum $\mathbf{k}$. At half filling, this model is gapped except at $m=0$, $m=2$, and $m=4$. In the gapped phases, the Chern number as a function of $m$ is
\begin{equation}
    C=\begin{cases}
        0& m<0,\\
        1& 0<m<2,\\
        -1& 2<m<4,\\
        0& m>4,
    \end{cases}
    \label{eq:masses}
\end{equation}
as depicted in Figure~\ref{fig: chern number plot}. 

Our focus for most of this work will be the transition at $m=0$, where the gap closes and the Chern number changes from $C=0$ to $C=1$. Near $m=0$, the low-energy, long wavelength physics is described by a single massless Dirac fermion with Lagrangian,
\begin{equation}
\label{eq: dirac lagrangian}
    \mathcal{L}=\bar{\psi}\,(i\hspace{.25 mm}\slashed{\partial}-m)\,\psi,
\end{equation}
where $\slashed{\partial}=\gamma^\mu \, \partial_\mu$ and $\bar{\psi}=\psi^\dagger \hspace{.25 mm} \gamma^0$. The two-component Dirac field $\psi$ describes modes of the electron in the band theory model. Our convention for Dirac matrices is
\begin{equation}
\label{eq: dirac matrices}
	\gamma^0=\sigma^1,\qquad \gamma^1=i  \hspace{.25 mm}\sigma^2,\qquad \gamma^2=i  \hspace{.25 mm}\sigma^3.
\end{equation}
Our primary goal in this work is to study this quantum critical point, described by Eq.~\eqref{eq: dirac lagrangian} with $m=0$, when the system has a boundary.

Before we introduce an edge to the system, we briefly review an important property of the bulk theory, Eq.~\eqref{eq: dirac lagrangian}, known as the parity anomaly~\cite{Deser-1982,Redlich1984a,Redlich1984b}. The mass term explicitly breaks parity and time-reversal symmetry, but even for $m=0$, the theory cannot be quantized in a time-reversal invariant way. 
The field theory of a massless Dirac fermion requires regularization, and it is common to introduce a Pauli-Villars fermion, adding a second fermion of large mass $\mu$ to regularize divergences. We started with a lattice model, Eq.~\eqref{eq:QHZ}, which is already a regularization, and in a local lattice model, there will necessarily be a second Dirac fermion~\cite{Nielsen1981},\footnote{There are actually three massive Dirac fermions at $m=0$ in Eq.~\eqref{eq:QHZ}, but two of them have opposite contributions to the Chern number.} which remains massive at $m=0$ and breaks time-reversal symmetry. We thus refer to the massive fermion introduced by the regularization as the heavy fermion doubler. If a background $U(1)$ gauge field\footnote{To be more precise, $A_\mu$ is a $U(1)$ spin$_c$ connection.} $A_\mu$ is coupled to both the Dirac fermion that becomes massless at the transition and the heavy fermion doubler, then the bulk critical theory may be described by the action,
\begin{equation}
\label{eq: dirac + background}
    S=\int d^3 r \, \, \bar{\psi}\, i\hspace{.2mm}\slashed{D}_A\, \psi-\frac{1}{8\pi}\int A\, dA,
\end{equation}
where we have integrated out the heavy fermion doubler, resulting in the Chern-Simons term. Here, we use the notation $A\, dA=d^3 x \,\, \varepsilon^{\mu\nu\lambda} \,A_\mu\, \partial_\nu A_\lambda$. The presence of this Chern-Simons term indicates that the critical theory has a fractional Hall conductivity,
\begin{equation}
    \sigma_{xy}=\frac{1}{2}\,\frac{e^2}{h},
\end{equation}
indicating that the effective Chern number at the transition is $C=1/2$. Thus, the bulk critical point does not have time-reversal symmetry, so if the system has a physical boundary, then we expect that there should be a signature of chirality at the edge even if there is no localized chiral edge mode that is sharply distinct from the critical bulk. Our goal in this work is to understand the physical consequences of the effective Chern number $C=1/2$.

Before proceeding, we should clarify that the level 1/2 Chern-Simons term in Eq.~\eqref{eq: dirac + background} is not completely correct since it is not gauge invariant globally. The more precise effect of the heavy fermion doubler is expressed in terms of the Atiyah-Patodi-Singer (APS) eta invariant of the Dirac operator~\cite{Atiyah1975,Alvarez-Gaume1985,Witten2016,Witten2021}. A proper regularization of the phase of the partition function for a massless Dirac fermion is
\begin{equation}
Z_\mathrm{Dirac}[A]=|\det(i\hspace{.2mm}\slashed{D}_A)|\,\exp\left(-\,\frac{i\hspace{.25mm}\pi}{2}\,\eta[A]\right),
\end{equation}
where the APS eta invariant is defined as
\begin{equation}
    \eta[A]=\lim_{\epsilon \, \to\, 0^+}\sum_n e^{-\,\epsilon \,(\lambda_n)^2}\sgn(\lambda_n)\, ,
\end{equation}
where $\lambda_n$ denote the eigenvalues of the Dirac operator $i\hspace{.25mm}\slashed{D}_A$, which are indexed by $n$. Thus, the eta invariant is a regularized version of the difference between the number of positive and negative eigenvalues of $i\hspace{.25mm}\slashed{D}_A$. The eta invariant is closely related to the Chern-Simons term. If the Dirac operator has no zero modes, then the variation of the eta invariant is related to the variation of the Chern-Simons term as\footnote{If the fermion is placed in a curved spacetime, a gravitational Chern-Simons term should be added to the Chern-Simons term for $A_\mu$.}
\begin{equation}
   -\, \frac{i}{2}\,\delta S_\mathrm{CS}[A]=-\,\frac{i\hspace{.25mm}\pi}{2}\,\delta \eta[A], \qquad S_\mathrm{CS}[A]=\frac{1}{4\pi}\int A\, dA,
\end{equation}
but the eta invariant itself is not equal to the Chern-Simons term globally.

Another way of phrasing the problem we address in this work is to ask how the eta invariant manifests itself in physical observables in the presence of a boundary. The eta invariant on manifolds with boundary has been studied in the mathematical physics literature~\cite{Muller1994}, and formal aspects of its role in anomaly inflow have been developed~\cite{Witten2021}. However, its direct consequences for physical observables at a quantum critical point with a boundary have not been fully explored.

\begin{figure}
		\centering
		\begin{subfigure}{0.46\textwidth}
			\includegraphics[width=\textwidth]{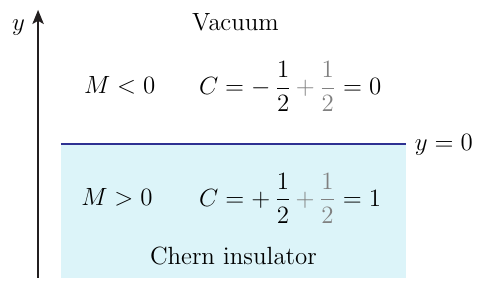}
			\caption{}
			\label{fig: chern insulator}
		\end{subfigure}
        \hspace{1cm}
		\begin{subfigure}{0.46\textwidth}
			\includegraphics[width=\textwidth]{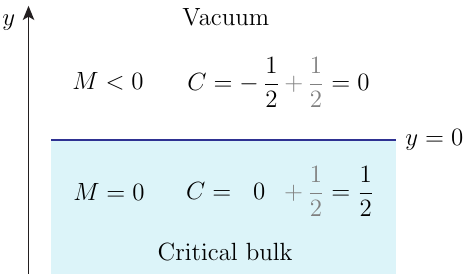}
			\caption{}
			\label{fig: critical bulk}
		\end{subfigure}
		\caption{Mass domain wall configurations. The $y>0$ region is regarded as vacuum with fixed Dirac mass $M<0$ while we tune to criticality in the material bulk, $y<0$, by a sign change in the Dirac mass. The heavy fermion doubler has a mass of fixed sign throughout space and contributes $1/2$ to the Chern number (depicted in gray) so that the $y>0$ region has vanishing Chern number. When the $y<0$ region has the same sign as the Dirac mass in the $y>0$ region, there is no protected edge mode, modeling a trivial insulator phase. \textbf{(a)} When the Dirac mass changes sign in space across $y=0$, there is a localized chiral edge mode, and the bulk is in its Chern insulator phase with Chern number $C=1$. \textbf{(b)} If the Dirac mass vanishes for $y<0$, then the bulk is at its critical point between the Chern insulator and the trivial insulator phase with an effective Chern number $C=1/2$.}
		\label{fig: mass domain walls}
	\end{figure}

Rather than working directly with the eta invariant, we use a physical construction with a fermion mass domain wall.\footnote{A similar discussion for a 3D domain wall can be found in Ref.~\cite{Boyanovsky1987}.} In the gapped Chern insulator phase, the edge modes are usually constructed by taking a domain wall across which the Dirac mass changes sign~\cite{Callan1985} as in Figure~\ref{fig: chern insulator}. In that case, one finds that a chiral fermion is exponentially localized along the interface of the wall. In contrast, if the fermion mass is the same on both sides of the wall, then there is no protected edge mode. We therefore model the Chern insulator transition using the configuration depicted in Figure~\ref{fig: critical bulk}. The region $y>0$ is treated as vacuum, with a fixed nonzero Dirac mass. The region $y<0$ is the material bulk, whose Dirac mass is tuned to zero at the Chern insulator transition. Indeed, if we fix the Dirac mass to be negative in the $y>0$ region, then the chiral edge mode at the interface disappears as the mass in the $y<0$ region continuously changes from positive to negative.

Meanwhile, the heavy fermion doubler has a mass of the same sign everywhere in space and also remains fixed across the Chern insulator transition. Its role is to properly quantize the Chern number in the gapped phases, but it plays essentially no direct role in the boundary physics. The only important constraint is that the heavy doubler must be chosen so that the net Hall response in the region $y>0$ vanishes, which ensures that the $y>0$ region can consistently be interpreted as vacuum, or as a trivial insulator.

This domain wall picture provides the setup for the continuum field theory problem we will treat below. Next, we will compute the fermion correlation function at the critical point with a boundary in this mass domain wall setup.

\section{Fermion correlation function}

\label{sec: two-point function}

\subsection{Step function mass domain wall}

\label{sec: step function mass}

We now specialize to the simplest realization of this domain wall setup that allows us to study the boundary physics at the Chern insulator critical point. The purpose of this construction is not to model microscopic details of a physical edge, but to isolate the universal long-distance physics of the critical boundary. On one side of the wall, the Dirac fermion remains gapless and describes the Chern insulator transition. On the other side, the fermion is given a large mass, so that this region acts as a gapped exterior. The interface at $y=0$ plays the role of the physical boundary of the critical system.

For concreteness, we take the mass profile to be a step function. This choice makes the problem exactly solvable while preserving the essential boundary physics. We take the Dirac mass as a function of the $y$ coordinate to be
\begin{equation}
    M(y)=-\,m_0\, \Theta(y)=\begin{cases}
        0, & y<0,\\
        -\, m_0,& y>0,
    \end{cases}
\end{equation}
where $\Theta(y)$ is the Heaviside step function and $m_0>0$ is a constant. We initially take $m_0$ to be finite, though we will eventually take the $m_0\to\infty$ limit. The action is
\begin{equation}
\label{eq: (2+1)d fermion action}
    S=\int d^3 r \, \bar{\psi}\left[i\hspace{.25 mm}\slashed{\partial}-M(y)\right]\psi.
\end{equation}
Our method for calculating the fermion two-point function will parallel the method of Ref.~\cite{Chandrasekharan1994}, which calculated the fermion propagator for a mass profile $M(y)=m_0 \tanh(m_0 y)$ to study anomaly inflow in the gapped Chern insulator phase. To determine the two-point function of the Dirac fermion $\psi$, we first solve the eigenvalue equation,
\begin{equation}
\label{eq: dirac eigenvalue eq}
\mathcal{D}\,\varphi_\lambda(t,x,y)\equiv\gamma^0\left[i\hspace{.25 mm}\slashed{\partial} -M(y)\right]\varphi_\lambda(t,x,y)=\lambda \, \varphi_\lambda(t,x,y).
\end{equation}
The eigenfunctions, $\varphi_\lambda(t,x,y)$, may then be used to determine the fermion two-point function as
\begin{equation}
\label{eq: propagator spectral}
    S_F(t-t',x-x';y,y')=\langle \mathcal{T}\,[\psi(t,x,y) \,\bar{\psi}(t',x',y')]\rangle =i\sum_\lambda \frac{\varphi_\lambda(t,x,y)\,\bar{\varphi}_\lambda(t',x',y')}{\lambda},
\end{equation}
where $\mathcal{T}$ denotes time-ordering, $\bar{\varphi}_\lambda=\varphi_\lambda^\dagger \gamma^0$, and we should introduce the Feynman $i \epsilon$ prescription in the formal sum over eigenvalues $\lambda$. The operator $\mathcal{D}$ is defined to be self-adjoint, implying that $\lambda\in\mathbb{R}$.

To solve the eigenvalue equation, Eq.~\eqref{eq: dirac eigenvalue eq}, we use translation invariance in $t$ and $x$, expressing the eigenfunctions of $\mathcal{D}$ in the form,
\begin{equation}
\label{eq: (2+1)d eigenmodes}
	\varphi_\lambda(t,x,y)=\frac{1}{2\pi}\,\phi_\lambda(y)\,e^{-ip_t t-ip_x x}, \qquad \phi_\lambda(y)=\begin{pmatrix}
	    \phi_{\lambda}^{(-)}(y)\\
        \phi_{\lambda}^{(+)}(y)
	\end{pmatrix}.
\end{equation}
In our basis of Dirac matrices, Eq.~\eqref{eq: dirac matrices}, the eigenvalue equation is
\begin{equation}
\label{eq: matrix eigenvalue eq}
	\begin{pmatrix}
		p_t -p_x & d/dy -M(y)\\ -d/dy-M(y) & p_t+p_x
	\end{pmatrix}	\begin{pmatrix}
	    \phi_{\lambda}^{(-)}(y)\\
        \phi_{\lambda}^{(+)}(y)
	\end{pmatrix}=\lambda\begin{pmatrix}
	    \phi_{\lambda}^{(-)}(y)\\
        \phi_{\lambda}^{(+)}(y)
	\end{pmatrix}.
\end{equation}
It is useful to parametrize the eigenvalues as
\begin{equation}
\label{eq: dirac eigenvalues}
	\lambda=p_t+\sigma \sqrt{p_x^2+q^2},
\end{equation}
where $\sigma=\pm 1$ and $q \geq 0$. We correspondingly write $\phi_{\lambda}(y)\to \phi_{\sigma, q}(y)$ to denote the solution of Eq.~\eqref{eq: matrix eigenvalue eq} with the eigenvalue in Eq.~\eqref{eq: dirac eigenvalues}. The form of the eigenvalues, Eq.~\eqref{eq: dirac eigenvalues}, may be determined by studying the asymptotics of Eq.~\eqref{eq: matrix eigenvalue eq}. For $y\to-\infty$, Eq.~\eqref{eq: matrix eigenvalue eq} reduces to the eigenvalue equation for a massless Dirac fermion, which has eigenvalues of the form Eq.~\eqref{eq: dirac eigenvalues} with $q$ representing the magnitude of the momentum of a scattering state in the $y$-direction. In the full domain wall problem, translation invariance in the $y$-direction is explicitly broken, but we may still parametrize the eigenvalues $\lambda$ in terms of the asymptotic momentum $q$.

By differentiating the coupled first-order differential equations in Eq.~\eqref{eq: matrix eigenvalue eq}, we may derive a decoupled set of second-order equations,
\begin{equation}
\label{eq: tise}
\left[-\frac{d^2 }{dy^2}\pm M'(y)+[M(y)]^2\right]\phi_{\sigma,q}^{(\pm)}(y)=q^2\, \phi_{\sigma,q}^{(\pm)}(y).
\end{equation}
Eq.~\eqref{eq: tise} is equivalent to the one-dimensional time-independent Schr\"{o}dinger equation with a ``potential'' of $V_\pm(y)=\pm \, M'(y)+[M(y)]^2=\mp \, m_0\, \delta(y)+m_0^2\,\Theta(y)$ and energy $q^2$. The solutions for $\phi_{\sigma,q}^{(\pm)}(y)$ may then be determined by solving the one-dimensional quantum mechanics problem, Eq.~\eqref{eq: tise}, and imposing that the solutions satisfy Eq.~\eqref{eq: matrix eigenvalue eq}.

Since we are mainly interested in the fermion two-point function, Eq.~\eqref{eq: propagator spectral}, in the low energy limit, we will ultimately take $m_0\to\infty$, so we only need to consider solutions to Eq.~\eqref{eq: matrix eigenvalue eq} for which $0 \leq  q < m_0$. We find that the solution for $y\leq 0$ is
\begin{equation}
\label{eq: scattering y<0}
	\phi_{\sigma,q}(y)=\frac{1}{2 m_0\sqrt{\pi \omega_q(\omega_q+\sigma\, p_x)}}\begin{pmatrix}
		iq \left[m_0\, e^{iqy}-\left(iq+\kappa_q\right) e^{-iqy}\right]\\
\left(p_x+\sigma\,\omega_q\right)\left[m_0\, e^{iqy}+\left(iq+\kappa_q\right) e^{-iqy}\right]
	\end{pmatrix},
\end{equation}
where $\omega_q=\sqrt{p_x^2+q^2}$ and $\kappa_q=\sqrt{m_0^2-q^2}$. For $y>0$, the solution is
\begin{equation}
\label{eq: scattering y>0}
	\phi_{\sigma,q}(y)=\frac{1}{2 m_0\sqrt{\pi \omega_q(\omega_q+\sigma\, p_x)}}e^{-\kappa_qy}\begin{pmatrix}
		iq\left( m_0-iq - \kappa_q\right)\\\left(p_x+\sigma \, \omega_q\right)\left(m_0+iq + \kappa_q\right)
	\end{pmatrix}.
\end{equation}
The solutions $\phi_{\sigma,q}(y)$ are normalized so that
\begin{equation}
\label{eq: normalization}
    \int_{-\infty}^\infty dy\, \phi_{\sigma,q}^\dagger(y)\,\phi_{\sigma', q'}(y)=\delta_{\sigma \sigma'}\,\delta(q-q'),
\end{equation}
which is the appropriate normalization for directly using Eq.~\eqref{eq: propagator spectral} to find the fermion propagator.

\subsection{Large \texorpdfstring{$m_0$}{m0} limit: Method of images}

\label{sec: method of images}

Having solved for the eigenfunctions of Eq.~\eqref{eq: dirac eigenvalue eq} for $0<q<m_0$, we are now equipped to use Eq.~\eqref{eq: propagator spectral} to determine the fermion propagator in the low energy limit, where $m_0\to\infty$. The fermion propagator for finite $m_0$ is determined in Appendix~\ref{sec: finite m0 propagator}. For $m_0\to\infty$, the $\phi_{\sigma,q}(y)$ vanish for $y>0$. But for $y<0$, on the side of the wall where the Dirac mass vanishes, we find
\begin{equation}
\label{eq: m large eigenfunctions}
	\phi_{\sigma, q}(y)=\frac{1}{\sqrt{\pi\omega_q(\omega_q+\sigma\, p_x)}} \begin{pmatrix}
		-\,q\sin(qy)\\ \left( p_x+\sigma\, \omega_q\right)\cos(q y)
	\end{pmatrix}.
\end{equation}
Using Eq.~\eqref{eq: propagator spectral}, we determine that the fermion two-point function with the Feynman $i\epsilon$ prescription is
\begin{align}
		S_F(t,x;y,y')&=i\sum_{\sigma=\pm 1} \int_0^\infty dq \int \frac{d^2 p}{(2\pi)^2}e^{-ip_t t-ip_x x} \phi_{\sigma, q}(y)\bar{\phi}_{\sigma, q}(y')\frac{1}{p_t+\sigma \sqrt{p_x^2+q^2}-i\hspace{.25mm}\sigma \hspace{.25mm} \epsilon} \\
&\hspace{-23mm}=i\int_{0}^{\infty}dq\int\frac{d^2{p}}{(2\pi)^2}\frac{2}{\pi}\frac{e^{-ip_t t-ip_x x}}{\omega_q^2-p_t^2-i\epsilon}\begin{pmatrix}
			-q\sin(qy)\cos(qy')&-(p_x+p_t)\sin(qy)\sin(qy')\\(p_x-p_t)\cos(qy)\cos(qy')&-q\cos(qy)\sin(qy')
		\end{pmatrix},\nonumber
        \end{align}
for $y\leq 0$ and $y'\leq 0$. To proceed, we perform a Wick rotation with $t\to -i \tau$ and $p_t\to i p_\tau$ to analytically continue to Euclidean signature. In our basis, the Dirac matrices in Euclidean signature are
\begin{equation}
    \Gamma^\tau=\gamma^0=\sigma^1, \qquad \Gamma^x=-\,i\hspace{.25mm}\gamma^1=\sigma^2, \qquad \Gamma^y=-\,i\hspace{.25mm}\gamma^2=\sigma^3.
\end{equation}
After calculating the integrals, we find that the fermion propagator in Euclidean signature is
\begin{equation}
\label{eq: images solution}
\langle \psi(r)\, \bar{\psi}(r')\rangle\equiv S_E(\tau-\tau',x-x';y,y')=S_0(r_-)+S_0(r_+)\, \Gamma^y, \qquad S_0(r)=\frac{\Gamma^\mu \, r_\mu}{4\pi |r|^3},
\end{equation}
where $(r_\pm)_\mu=(\tau-\tau',x-x',y\pm y')$ and $S_0$ is the bulk propagator for a massless Dirac fermion in (2+1)d Euclidean spacetime.

Another way to determine the fermion propagator without directly calculating the integrals is to observe that the solutions in the $m_0\to\infty$ limit, Eq.~\eqref{eq: m large eigenfunctions}, satisfy the boundary condition,
\begin{equation}
-\Gamma^y\, \phi_{\sigma,q}(y=0)=\phi_{\sigma,q}(y=0), \qquad \bar{\phi}_{\sigma,q}(y=0)\, (-\Gamma^y)=-\,\bar{\phi}_{\sigma,q}(y=0).
\end{equation}
By Eq.~\eqref{eq: propagator spectral}, this boundary condition on the eigenfunctions of $\mathcal{D}$ implies that the propagator in Euclidean signature obeys the boundary conditions,
\begin{align}
\label{eq: propagator bc}
\begin{split}
    (-\Gamma^y)\, S_E(\tau,x;0,y') &=S_E(\tau,x;0,y'), \\
    S_E(\tau,x;y,0)\, (-\Gamma^y)&=-\,S_E(\tau,x;y,0).
    \end{split}
\end{align}
Thus, in the $m_0\to\infty$ limit, the fermion propagator satisfies
\begin{equation}
    \Gamma^\mu  \partial_\mu \, S_E(\tau-\tau',x-x';y,y')=\delta^{(3)}(r-r'),
\end{equation}
in the region $y<0$ and $y'<0$ subject to the boundary conditions in Eq.~\eqref{eq: propagator bc}.

Hence, in the $m_0\to\infty$ limit, the problem is equivalent to that of a massless Dirac fermion $\psi(\tau,x,y)$ in a half-space that obeys the boundary condition,
\begin{equation}
\label{eq: fermion bc}
    -\Gamma^y\, \psi(\tau,x,y=0)= \psi(\tau,x,y=0), \qquad \bar{\psi}(\tau,x,y=0)(-\Gamma^y)= -\,\bar{\psi}(\tau,x,y=0).
\end{equation}
This problem may be obtained using the method of images. The action for a massless Dirac fermion is invariant under a reflection symmetry,
\begin{equation}
\label{eq: reflection symmetry}
    \psi(\tau,x,y)\to \zeta_y\,\Gamma^y \psi(\tau,x,-y),
\end{equation}
where $\zeta_y\in U(1)$ is a phase factor. If we choose $\zeta_y=-1$, then the boundary condition in Eq.~\eqref{eq: fermion bc} is equivalent to imposing that $\psi$ is invariant at $y=0$ under a reflection, $y\to -y$. In Eq.~\eqref{eq: images solution}, the second term,
\begin{equation}
    S_0(r_+)\Gamma^y=S_0(\tau-\tau',x-x',y-(-y'))\Gamma^y,
\end{equation}
is interpreted as the image obtained by reflecting the point $(\tau',x',y')\to (\tau',x',-y')$. The image solution, Eq.~\eqref{eq: images solution}, then satisfies
\begin{equation}
    \Gamma^\mu \partial_\mu \, S_E(\tau-\tau',x-x';y,y')=\mathbb{I}_2 \, \delta^{(3)}(r_-)+\Gamma^y \,\delta^{(3)}(r_+),
\end{equation}
and the second delta function term vanishes if $y<0$ and $y'<0$. Eq.~\eqref{eq: images solution} also satisfies the boundary conditions, Eq.~\eqref{eq: propagator bc}.

Other boundary conditions besides those in Eq.~\eqref{eq: fermion bc} are also allowed. For the Hamiltonian to be self-adjoint, the boundary condition should ensure that the fermion current normal to the boundary vanishes~\cite{McCann2004,Akhmerov2008,Shtanko2018,Biswas2022},
\begin{equation}
    J^y(\tau,x,y=0)=-\, i \hspace{.25mm}\bar{\psi}\, \Gamma^y \, \psi\,|_{y=0}=0,
\end{equation}
which is satisfied by Eq.~\eqref{eq: fermion bc}. However, the only boundary conditions consistent with the residual conformal symmetry parallel to the boundary are~\cite{Biswas2022}
\begin{equation}
\label{eq: fermion bc both chiralities}
    -\Gamma^y\, \psi(\tau,x,y=0)= \pm\,\psi(\tau,x,y=0), \qquad \bar{\psi}(\tau,x,y=0)(-\Gamma^y)= \mp\,\bar{\psi}(\tau,x,y=0).
\end{equation}
For both of these boundary conditions, the fermion propagator may be solved using the method of images,
\begin{equation}
\label{eq: fermion correlations both bc}
    S_\pm(r,r')=S_0(r_-)\pm S_0(r_+)\,\Gamma^y, \qquad S_\pm(r,r')\equiv S_\pm(\tau-\tau',x-x';y,y').
\end{equation}
If we had taken a Dirac mass of the opposite sign for $y>0$, then we would have obtained the boundary condition $-\Gamma^y\, \psi(\tau,x,y=0)=-\psi(\tau,x,y=0)$ after taking the magnitude of this mass to infinity.

\subsection{Boundary correlation function}

\label{sec: boundary propagator}

Now that we have the fermion two-point function in the presence of a boundary, Eq.~\eqref{eq: images solution}, we are prepared to examine the fermion correlation function at the edge when the bulk is at its quantum critical point. For $y=y'=0$, the fermion correlation function is
\begin{equation}
\label{eq: boundary propagator}
	S_E(\tau,x;0,0)=\frac{1}{2\pi}\frac{\Gamma^a \, \rho_a}{|\rho|^3}\frac{\mathbb{I}_2+ \Gamma^y}{2}=\frac{\Gamma^+}{4\pi \,|\rho| \, x^+},
\end{equation}
where $\rho_a=(\tau,x)$, $|\rho|=\sqrt{\tau^2+x^2}=\sqrt{2x^+x^-}$, $\Gamma^\pm=(\Gamma^x\pm i \hspace{.25mm}\Gamma^\tau)/\sqrt{2}$, and $x^\pm=(x\pm i\hspace{.25mm} \tau)/\sqrt{2}$. In Lorentzian signature, this boundary correlation function is expressed as Eq.~\eqref{eq: bdry propagator intro}. For comparison, in the gapped Chern insulator phase with $C=1$, the electron two-point function at the edge has the typical chiral fermion correlation function,
\begin{equation}
\label{eq: chiral fermion}
    \langle \psi(\tau,x,y=0)\, \bar{\psi}(0)\rangle\sim \frac{\Gamma^+}{x^+}.
\end{equation}
By comparing with Eq.~\eqref{eq: boundary propagator}, we see that the critical boundary correlation carries a similar chiral structure but is dressed by the factor $1/|\rho|$. The boundary propagator in Eq.~\eqref{eq: boundary propagator} is that of a (1+1)d chiral fermion that has acquired an anomalous dimension through dressing by the critical bulk degrees of freedom.

Next, we clarify in what sense the critical boundary correlation function in Eq.~\eqref{eq: boundary propagator} is chiral. As discussed previously, the fermion obeys the boundary condition in Eq.~\eqref{eq: fermion bc}. In the effective (1+1)d boundary theory, eigenspinors of $\Gamma^5=-\Gamma^y=i\hspace{.25mm}\Gamma^\tau\,\Gamma^x$ transform in a chiral spin-$1/2$ representation under rotations about the $y$-axis (or Lorentz boosts parallel to the boundary in Lorentzian signature). Indeed, the operator that rotates a boundary spinor by angle $\theta$ is given by
\begin{equation}
    R(\theta)=\exp\left(-\,\frac{\theta}{2}\,\Gamma^\tau\,\Gamma^x\right)=\exp\left(\frac{i}{2}\,\theta \,\Gamma^5\right).
\end{equation}
Hence, because the boundary propagator in Eq.~\eqref{eq: boundary propagator} satisfies the boundary condition in Eq.~\eqref{eq: fermion bc}, at the boundary the fermion transforms as a chiral fermion under rotations. On the other hand, unlike Eq.~\eqref{eq: chiral fermion}, because of the coupling to the gapless bulk, Eq.~\eqref{eq: boundary propagator} depends on both $x^+$ and $x^-$ since we may write
\begin{equation}
\label{eq: chiral dependence}
    S_E(\tau,x;0,0)\sim \frac{\Gamma^+}{(x^+)^{3/2}(x^-)^{1/2}}.
\end{equation}
However, the dependence on $x^+$ and $x^-$ is asymmetric since these coordinates carry different powers in Eq.~\eqref{eq: chiral dependence}. The final sense in which the boundary is chiral, which we will discuss in Section~\ref{sec: chiral currents}, is that the structure of the currents at the boundary is purely chiral. This unusual chiral boundary correlation function and its related observables are direct physical consequences of the effective Chern number $C=1/2$ that characterizes the bulk quantum critical point.

\subsection{Approaching the bulk critical point}

\label{sec: chern insulator near criticality}

Having established the nature of the fermion correlation function when the bulk sits exactly at the critical point between a Chern insulator with Chern number $C=1$ and a trivial insulator, we now examine the behavior of the fermion two-point function as the critical point is approached from the $C=1$ phase. We will then be able to study exactly how the chiral edge mode delocalizes as the bulk approaches criticality.

We now consider a Dirac fermion with the Lagrangian in Eq.~\eqref{eq: (2+1)d fermion action} but with a mass profile, 
\begin{equation}
\label{eq: mass sign change}
    M(y)=-\,m_0 \, \Theta(y)+m\, \Theta(-y)=\begin{cases}
        m & y<0,\\
        -\, m_0 & y>0.
    \end{cases}
\end{equation}
Indeed, for fixed $m_0>0$, there will be a localized chiral edge mode for $m>0$, but the edge mode will not exist for $m<0$, so we may model the Chern insulator near a transition to a trivial insulator by taking $m_0 \gg m>0$.

Since the calculations are similar to those in Sections~\ref{sec: step function mass} and \ref{sec: method of images}, we present the derivation in Appendix~\ref{sec: near criticality propagator calc} and simply state the result here. The fermion propagator in Euclidean signature in the $m_0\to\infty$ limit with finite $m>0$ is
\begin{align}
\label{eq: finite m propagator}
\begin{split}
S_E(r,r')&=S_m(r_-)+S_m(r_+)\,\Gamma^y-\frac{m \, e^{-m|r_+|}}{4\pi |r_+|}(\mathbb{I}_2+\Gamma^y)\\
&\hspace{5mm}+m\left[e^{-m |r_+|}\left(\frac{y+y'}{|r_+|}-1\right)+2\,e^{-m|y+y'|}\right] \frac{\Gamma^+}{4\pi \,(x^+-x'^+)},
\end{split}
\end{align}
where $S_m$ is the propagator for a fermion of mass $m$ everywhere in (2+1)d Euclidean spacetime,
\begin{equation}
    S_m(r)=(-\,\Gamma^\mu \,\partial_\mu +m)\,\frac{e^{-m |r|}}{4\pi |r|}=\frac{e^{-m |r|}}{4\pi |r|^3}\left[\Gamma^\mu\, r_\mu (1+m \,|r|)+m \,|r|^2\right].
\end{equation}
Unlike in the $m=0$ case treated in Section~\ref{sec: method of images}, a solution by the method of images is not possible for $m>0$. While the naive solution by the method of images, 
\begin{equation}
\label{eq: naive solution}
S_\mathrm{naive}(r,r')=S_m(r_-)+S_m(r_+)\,\Gamma^y,
\end{equation}
indeed satisfies the differential equation for the propagator,
\begin{equation}
    (\Gamma^\mu \, \partial_\mu \, +m)\, S_\mathrm{naive}(r,r')=\delta^{(3)}(r-r'),
\end{equation}
in the region where $y<0$ and $y'<0$, it does not satisfy the boundary condition, Eq.~\eqref{eq: propagator bc}. The physical reason why the method of images does not work here is that a massive Dirac fermion is not invariant under the reflection symmetry in Eq.~\eqref{eq: reflection symmetry}. Thus, it is necessary to solve for the propagator by summing over the eigenmodes as in Appendix~\ref{sec: near criticality propagator calc}.

The last term in Eq.~\eqref{eq: finite m propagator} represents the chiral mode, which is exponentially localized at the edge because of the factor $e^{-m|y+y'|}$. For small $m>0$, expanding Eq.~\eqref{eq: finite m propagator} to linear order in $m$ gives
\begin{align}
\begin{split}
	S_E(r,r')&= \frac{\Gamma^\mu (r_-)_\mu}{4\pi |r_-|^3}+\frac{\Gamma^\mu (r_+)_\mu}{4\pi |r_+|^3}\,\Gamma^y+\frac{m}{4\pi}\left(\frac{1}{|r_-|}-\frac{1}{|r_+|}\right)\mathbb{I}_2+m\left(\frac{y+y'}{|r_+|}+1\right)\frac{\Gamma^+}{4\pi \,(x^+-x'^+)}\\
    &\hspace{5mm}+\mathcal{O}(m^2),
    \end{split}
\end{align}
so we see that the chiral mode delocalizes from the edge, leaking into the bulk as the Chern insulator approaches its quantum critical point. At the boundary, $y=y'=0$, Eq.~\eqref{eq: finite m propagator} becomes
\begin{equation}
	S_E(r,r'=0)|_{y=0}=\left(\frac{e^{-m |\rho|}}{|\rho|}+2\hspace{.25mm}m\right)\frac{\Gamma^+}{4\pi \, x^+}=\left(\frac{1}{|\rho|}+m\right)\frac{\Gamma^+}{4\pi x^+}+\mathcal{O}(m^2),
\end{equation}
where $|\rho|=\sqrt{\tau^2+x^2}$ is the boundary separation. Thus, for $m> 0$, the chiral fermion of the gapped Chern insulator phase is the dominant contribution at long distances. On the other hand, at $m=0$, we see that we recover the boundary correlation function in Eq.~\eqref{eq: boundary propagator}, which resembles a chiral fermion that has acquired an anomalous dimension as discussed in Section~\ref{sec: boundary propagator}.

\section{Charge currents and anomaly inflow}

\label{sec: currents and anomalies}

\subsection{Current correlation functions}

\label{sec: current correlation functions}

In addition to the fermion correlation function, other important observables include the correlation functions of the fermion current, $J^\mu=-\,i\, \bar{\psi}\,\Gamma^\mu\, \psi$. Before directly calculating this observable for the gapless free Dirac fermion, we explain generic constraints from the residual conformal invariance, generalizing the discussion in Ref.~\cite{McAvity1995} for (2+1)d CFTs in a half-space that lack time-reversal symmetry. It is useful to define
\begin{gather}
    \label{eq: I tensor}
    I_{\mu\nu}(r_-)=\delta_{\mu\nu}-\frac{2\, (r_-)_\mu \, (r_-)_\nu}{|r_-|^2}, \\
    X_\mu= v\left(\frac{2\hspace{.25mm}y}{|r_-|^2}\,(r_-)_\mu-\delta_{\mu y}\right), \qquad  X'_\mu =-\,v\left(\frac{2\hspace{.25mm}y'}{|r_-|^2}\,(r_-)_\mu+\delta_{\mu y}\right),\label{eq: X vectors}
\end{gather}
where $(r_\pm)_\mu=(\tau-\tau',x-x',y\pm y')$ and $v=|r_-|/|r_+|$, because these objects transform covariantly under a conformal transformation $R_{\mu \alpha}(r)$ that preserves the boundary~\cite{McAvity1995},
\begin{equation}
    I_{\mu\nu}(r_-)\to R_{\mu\alpha}(r)\, R_{\nu\beta}(r')\, I^{\alpha\beta}(r_-),\qquad X_\mu\to R_{\mu\nu}(r)\, X^\nu,\qquad X'_\mu\to  R_{\mu\nu}(r')\, (X')^\nu.
\end{equation}
For a (2+1)d CFT in a half-space, the most general two-point function for a conserved $U(1)$ current $J_\mu$ is
\begin{align}
\label{eq: vector 2 pt func}
\begin{split}
    \langle J_\mu(r)\, J_\nu(r')\rangle &=\frac{1}{|r_-|^4}\left[I
    _{\mu\nu}(r_-)\,F(v)+X_\mu \, X'_\nu \,G(v)+i\,\varepsilon_{\mu\lambda\sigma} \, X^\lambda \,{I^{\sigma}}_\nu(r_-)\,H(v)\right]\\
    &\hspace{5mm}+\frac{i\hspace{.25mm}\kappa}{2\pi}\, \varepsilon_{\mu\nu\lambda}\, \partial^\lambda\, \delta^{(3)}(r_-),
    \end{split}
\end{align}
for some functions $F$, $G$, and $H$ and a constant $\kappa$. Under parity and time-reversal, the first two terms are even while the last two terms are odd. The first two terms were described in Refs.~\cite{McAvity1995,Herzog2017}. The third term was previously noticed in Ref.~\cite{Wang2021}, but the relation of $H(v)$ to anomalies was not worked out in full generality, as we will do here. Physically, this term represents a chiral mode that delocalizes into the bulk, decaying away from the edge with a power law. We also note that $\varepsilon_{\mu\lambda\sigma} \, (X')^\lambda \,{I^{\sigma}}_\nu(r_-)=\varepsilon_{\nu\lambda\sigma} \, X^\lambda \,{I^{\sigma}}_\mu(r_-)$ is not an independent term.

Next, we discuss general constraints on $F$, $G$, and $H$. Conservation of $J_\mu$,
\begin{equation}
    \partial^\mu \langle J_\mu(r)\, J_\nu(r')\rangle = 0,
\end{equation}
for $y<0$ and $y'<0$ imposes that~\cite{McAvity1995}
\begin{equation}
\label{eq: F, G constraint}
    	v\,\frac{d}{dv}[F(v)+G(v)]=2\,G(v).
\end{equation}
Deep in the bulk, where $y\to-\infty$ and $y'\to-\infty$ with $y-y'$ held fixed, Eq.~\eqref{eq: F, G constraint} implies that $G(0)=0$, assuming $F$ and $G$ are smooth. Taking the deep bulk limit of Eq.~\eqref{eq: vector 2 pt func} then gives
\begin{equation}
    \langle J_\mu(r)\, J_\nu(r')\rangle_{\mathrm{bulk}}=\frac{1}{|r_-|^4}\left[I
    _{\mu\nu}(r_-)\,F(0)+i\,\varepsilon_{\mu\nu\lambda} \, \frac{r_-^\lambda}{|r_-|}\,H(0)\right]+\frac{i\hspace{.25mm}\kappa}{2\pi}\, \varepsilon_{\mu\nu\lambda}\, \partial^\lambda\, \delta^{(3)}(r_-),
\end{equation}
but the second term is not compatible with full conformal invariance in the deep bulk, so we require $H(0)=0$. In contrast, $F(0)$ and $\kappa$ are generically nonzero and are fixed by the bulk theory.

As we will discuss in Section~\ref{sec: anomaly inflow}, the anomalies are encoded in the time-reversal odd terms in
\begin{equation}
    \langle J_y(r)\, J_\nu(r')\rangle |_{\,y\,\to\, 0^-}.
\end{equation}
For completeness, we note that the time-reversal even terms are
\begin{align}
\label{eq: even part boundary}
\begin{split}
    \langle J_y(r)\, J_y(r')\rangle_{\mathrm{even}}|_{y\to 0^-} &=[F(1)+G(1)]\,\frac{|\rho-\rho'|^2-(y')^2}{[|\rho-\rho'|^2+(y')^2]^3},\\
    \langle J_y(r)\, J_a(r')\rangle_{\mathrm{even}}|_{y\to 0^-}& =[F(1)+G(1)]\,\frac{2\, y' \, (\rho_a-\rho'_a)}{[|\rho-\rho'|^2+(y')^2]^3},
    \end{split}
\end{align}
where we recall that $a\in \lbrace \tau, x\rbrace$, $\rho_a=(\tau,x)$, and $|\rho|=\sqrt{\tau^2+x^2}$. The time-reversal odd parts are
\begin{equation}
\label{eq: odd boundary current}
   \langle J_y(r)\, J_\nu(r')\rangle_{\mathrm{odd}}|_{y\to 0^-}=\frac{i\hspace{.25mm}(-\,k_H+\kappa)}{4\pi}\, \varepsilon_{y\nu a}\,\partial^a \delta^{(2)}(\rho-\rho')\, \delta_-(y'),
\end{equation}
where $\delta_-(y)$ is defined so that
\begin{equation}
    \int_{-\infty}^0 dy\, \Phi(y)\,\delta_-(y)=\Phi(0)
\end{equation}
for a function $\Phi(y)$. As demonstrated in Appendix~\ref{sec: delocalized anomaly}, the coefficient $k_H$ is
\begin{equation}
\label{eq: anomaly coefficient main}
    k_H=2\pi^2\int_0^1 \frac{dv}{v^4} \left[(1+v^2)\,\operatorname{arctanh}(v)-v\right]H(v).
\end{equation}
As we will explain in more detail in Section~\ref{sec: anomaly inflow}, this coefficient characterizes the anomaly of the delocalized chiral mode.

Finally, we now specialize to the free massless Dirac fermion. The two-point function of the current $J^\mu=-\,i\, \bar{\psi}\,\Gamma^\mu\, \psi$ is
\begin{equation}
\label{eq: cft chern transition currents}
    \Pi^{\mu\nu}(r,r')\equiv\langle J^\mu(r) \, J^\nu(r')\rangle =\Tr\left[ \Gamma^\mu S_\pm(r,r') \Gamma^\nu S_\pm(r',r)\right],
\end{equation}
where we use Eq.~\eqref{eq: fermion correlations both bc} for the fermion propagator, which allows us to treat both boundary conditions in Eq.~\eqref{eq: fermion bc both chiralities}. By a direct calculation, we may confirm that $\Pi^{\mu\nu}$ has the form given in Eq.~\eqref{eq: vector 2 pt func} with
\begin{equation}
\label{eq: FGH functions}
	F(v)=\frac{1}{8\pi^2}\left(1+v^4\right), \qquad G(v)=-\,\frac{1}{4\pi^2}\,v^4,\qquad H(v)=\pm \,\frac{1}{4\pi^2}\, v^2,\qquad \kappa=0,
\end{equation}
which indeed satisfy Eq.~\eqref{eq: F, G constraint} and $G(0)=H(0)=0$. We thus have
\begin{equation}
\label{eq: div pi}
    \partial_\mu \Pi^{\mu\nu}(r,r')=0,
\end{equation}
for $y<0$ and $y'<0$. We also have $F(1)+G(1)=0$ so that Eq.~\eqref{eq: even part boundary} vanishes, and Eq.~\eqref{eq: anomaly coefficient main} gives $k_H=\pm\, 1/2$. By Eq.~\eqref{eq: odd boundary current}, we find
\begin{equation}
\label{eq: boundary polarization tensor}
\langle J_y(r)\, J_\nu(r')\rangle |_{y\to 0^-}=\mp \, \frac{i}{8\pi}\,\varepsilon^{y\nu a}\,\partial_a \,\delta^{(2)}(\rho-\rho')\,\delta_-(y'),
\end{equation}
which will also be important for analyzing anomalies in Section~\ref{sec: anomaly inflow}. Now that we have the current two-point functions, we can analyze their physical meaning and then use them to understand anomaly matching at the quantum critical point.


\subsection{Chiral currents}

\label{sec: chiral currents}

To elucidate the physical signatures of the delocalized chiral modes more transparently, it is useful to define the chiral currents, $J_\mp=J^\pm=(J^x\pm i \hspace{.25mm}J^\tau)/\sqrt{2}$. We take the fermion to obey the right-moving boundary condition in Eq.~\eqref{eq: fermion bc}.

Then, when the bulk is at its critical point between the Chern insulator and the trivial insulator, the two-point functions of the chiral currents are
\begin{equation}
\label{eq: chiral currents}
    \langle J_\pm(r)\,J_\pm(r')\rangle =-\,\frac{(x^\mp-x'^\mp)^2}{4\pi^2}\left(\frac{1}{|r_-|^3}\pm \frac{1}{|r_+|^3}\right)^2.
\end{equation}
In the basis of $J_\pm$ and $J_y$, the only current two-point function that does not vanish at the boundary of the system is
\begin{equation}
\label{eq: chiral boundary current}
    \langle J_+(r)\,J_+(0)\rangle|_{y=0}=-\,\frac{1}{4\pi^2}\,\frac{1}{|\rho|^2 }\,\frac{1}{(x^+)^2},
\end{equation}
which is another signal of chirality at the edge. This chiral mode leaks into the bulk but decays with a power law. Meanwhile, the two-point function for the current of the opposite chirality, $J_-$, vanishes at the boundary and grows in the bulk. Deep in the bulk, the two-point functions of both chiralities become comparable. These observations also clarify the relation of the delocalized chiral mode to the edge. Because the boundary is no longer decoupled from the bulk, one might argue that the delocalized chiral mode is not really an edge mode. However, its existence is only possible because the system has an edge, and as we have observed, the delocalized chiral mode indeed disappears deep in the bulk.

\begin{figure}

\centering\includegraphics[width=.75\textwidth]{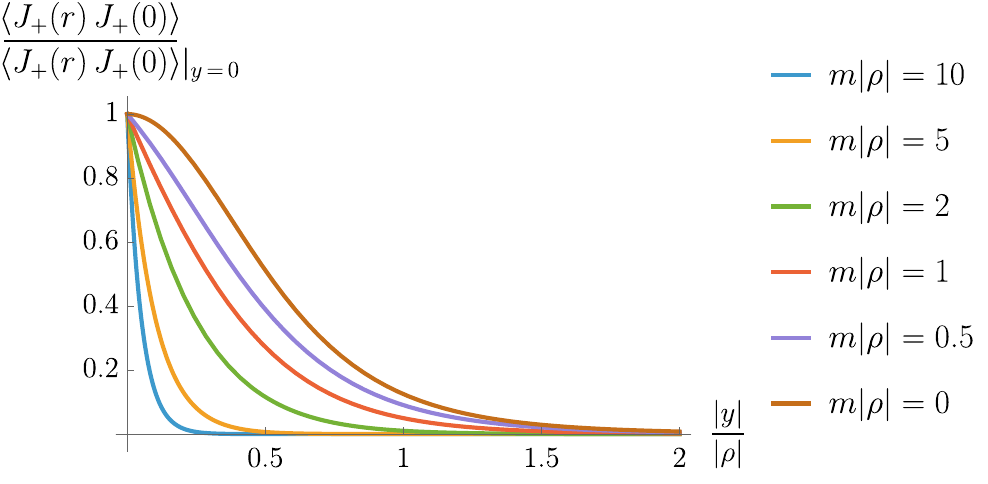}
	\caption{For a generic point $r=(\tau,x,y)$ and $r'=0$, we plot the ratio of current correlation functions $\langle J_+(r)\, J_+(0) \rangle/ \langle J_+(r)\, J_+(0)\rangle|_{y=0}$ from Eqs.~\eqref{eq: bdry j+} and \eqref{eq: j+ correlation} as a function of $|y|/|\rho|$ for various values of $m |\rho|\geq 0$. We recall that $|y|$ denotes the distance from the boundary while $|\rho|=\sqrt{\tau^2+x^2}$ is the distance parallel to the boundary from the origin. For $m>0$, this profile decays exponentially with a length scale set by $1/m$, but for $m=0$, we obtain power law decay.}
	\label{fig: current plot}
\end{figure}

Next, we consider the case in which the bulk is in its gapped Chern insulator phase. We may then use the fermion propagator, Eq.~\eqref{eq: finite m propagator}, at finite $m>0$. Again, the only nonzero current two-point function at the boundary is
\begin{equation}
\label{eq: bdry j+}
    \langle J_+(r)\,J_+(0)\rangle|_{y=0} = -\,\frac{1}{4\pi^2}\left[\frac{e^{-m|\rho|}\,(1+m\,|\rho|)}{|\rho|}+m\,(2-e^{-m|\rho|})\right]^2 \frac{1}{(x^+)^2}.
\end{equation}
At long distances, this current correlation function depends on $x^+$ and goes as $\sim 1/(x^+)^2$. As depicted in Figure~\ref{fig: current plot}, this current two-point function decays exponentially into the bulk with a length scale set by $1/m$,
\begin{align}
\label{eq: j+ correlation}
\begin{split}
    \langle J_+(r)\, J_+(r')\rangle &=-\,\frac{1}{4\pi^2}\left\lbrace (x^--x'^-)\left(\frac{e^{-m |r_-|}(1+m\, |r_-|)}{|r_-|^3}+\frac{e^{-m |r_+|}(1+m \,|r_+|)}{|r_+|^3}\right)\right.\\
    &\hspace{5mm}\left.+\frac{m}{x^+-x'^+}\left[e^{-m |r_+|}\left(\frac{y+y'}{|r_+|}-1\right)+2\,e^{-m|y+y'|}\right]\right\rbrace^2,
    \end{split}
\end{align}
signaling that the chiral mode is localized at the edge in the Chern insulator phase, but as $m\to 0$ the chiral current delocalizes. Meanwhile, the two-point function for the current of the opposite chirality is
\begin{align}
    \langle J_-(r)\,J_-(r')\rangle =-\,\frac{(x^+-x'^+)^2}{4\pi^2}\left[\frac{e^{-m |r_-|}(1+m \,|r_-|)}{|r_-|^3}-\frac{e^{-m |r_+|}(1+m \,|r_+|)}{|r_+|^3}\right]^2,
\end{align}
which not only vanishes at the boundary but also decays exponentially in the bulk.

Following the analysis in Section~\ref{sec: chern insulator near criticality}, we can also calculate the two-point function of the right-moving current $J_+$ to quadratic order in $m$, which gives
\begin{equation}
   \langle J_+(r)\, J_+(0)\rangle|_{y=0} =-\,\frac{1}{4\pi^2}\left(\frac{1}{|\rho|^2}+\frac{2\hspace{.25mm}m}{|\rho|}+2\hspace{.25mm}m^2\right)\frac{1}{(x^+)^2}+\mathcal{O}(m^3).
\end{equation}
Thus, as the bulk critical point is approached, the mass-dependent contribution from the localized chiral edge mode disappears, and the current correlation function reduces to Eq.~\eqref{eq: chiral boundary current}.

\subsection{Anomaly inflow at criticality}

\label{sec: anomaly inflow}

Another important application of the current correlation functions derived in Section~\ref{sec: current correlation functions} is the study of anomaly inflow at the quantum critical point. Anomaly matching has previously been studied in momentum space~\cite{Prochazka2019}, in contrast to our position-space approach here. We couple the Dirac fermion to the background electromagnetic field $A_\mu$. We integrate out the fermion and expand the functional determinant to one-loop order, which leads to an effective action for $A_\mu$,
\begin{equation}
\label{eq: effective action}
	S_\mathrm{eff}[A_\mu]=-\,\frac{1}{2}\int_{y\, <\, 0} d^3 r \int_{y'\, <\, 0} d^3 r' \,A_\mu(r)\,\Pi^{\mu\nu}(r,r') \, A_\nu(r')-\frac{i}{8\pi}\int_{y\, >\, 0} A\, dA, 
\end{equation}
with $\Pi^{\mu\nu}(r,r')\equiv \langle J^\mu(r)\, J^\nu(r')\rangle$ given by Eq.~\eqref{eq: cft chern transition currents}. Until now, we have neglected the effects of the fermion propagator for $y>0$ since it vanishes in the $m_0\to\infty$ limit. However, a proper analysis should calculate the response to $A_\mu$ at finite $m_0$ first and then take $m_0\to\infty$. As demonstrated in Appendix~\ref{sec: finite m0 propagator}, this leads to the Chern-Simons response in Eq.~\eqref{eq: effective action} for $y>0$.

As discussed in Section~\ref{sec: physical setup}, we should also include a response for the heavy fermion doubler. Since we regard the $y>0$ region as a trivial insulator, we should choose the heavy fermion doubler so that its Chern-Simons response cancels the response for $y>0$ in Eq.~\eqref{eq: effective action}. Thus, the response of the heavy fermion doubler is
\begin{equation}
\label{eq: PV response}
    S_\mathrm{PV}[A_\mu]=\frac{i}{8\pi}\int_{\mathbb{R}^3}A\, dA,
\end{equation}
and this action has support everywhere in spacetime since the heavy fermion doubler has the same mass everywhere.

Adding Eq.~\eqref{eq: effective action} and Eq.~\eqref{eq: PV response} together gives a full effective action for $A_\mu$ given by
\begin{align}
\label{eq: response}
\begin{split}
    S_{\mathrm{resp}}[A_\mu]&=S_{\mathrm{matter}}[A_\mu]+S_{\mathrm{CS}}[A_\mu],\\
    S_{\mathrm{matter}}[A_\mu]&=-\,\frac{1}{2}\int_{\mathbb{R}^3} d^3 r \int_{y'\, <\, 0} d^3 r'\, A_\mu(r) \,K^{\mu\nu}(r,r')\, A_\nu(r'),\\
    S_{\mathrm{CS}}[A_\mu]&=\frac{i}{8\pi}\int_{y\, <\, 0} A\, dA.
    \end{split}
\end{align}
where we define the kernel,
\begin{equation}
\label{eq: K kernel}
    K^{\mu\nu}(r,r')=\Theta(-y)\,\Pi^{\mu\nu}(r,r'),
\end{equation}
where $\Theta$ denotes the Heaviside step function. We are now prepared to study anomaly matching for Eq.~\eqref{eq: response}. Under a gauge transformation, $A_\mu \to A_\mu +\partial_\mu \xi$, the Chern-Simons term changes by
\begin{equation}
\label{eq: level 1/2 CS gauge trans}
    \Delta S_{\mathrm{CS}}=\frac{i}{8\pi}\int_{y\, =\, 0}d^2 \rho \, \xi \, F_{\tau x},
\end{equation}
where $F_{\tau x}=\partial_\tau A_x-\partial_x A_\tau=\varepsilon^{ab}\, \partial_a A_b$. Meanwhile, the response to the gapless Dirac fermion changes by
\begin{equation}
		\Delta S_\mathrm{matter}=-\frac{1}{2}\int_{\mathbb{R}^3} d^3 r \int_{y'\, <\, 0} d^3 r'\, \partial_\mu\xi(r) \,K^{\mu\nu}(r,r')\, [2\, A_\nu(r')+\partial_\nu'\xi(r')].
\end{equation}
Since $K^{\mu\nu}(r,r')$ vanishes at infinity, we may integrate by parts to obtain
\begin{equation}
\label{eq: matter resp gauge trans}
	\begin{split}
    \Delta S_\mathrm{matter}=\frac{1}{2}\int_{\mathbb{R}^3} d^3 r \int_{y'\, <\, 0} d^3 r'\, \xi(r) \,[\partial_\mu K^{\mu\nu}(r,r')]\, [2\, A_\nu(r')+\partial_\nu'\xi(r')],
	\end{split}
\end{equation}
In the region where $y<0$ and $y'<0$, we recall that $\partial_\mu \Pi^{\mu\nu}=0$, which implies that $\partial_\mu K^{\mu\nu}=0$ for $y<0$ and $y'<0$. However, because of the step functions in the definition of $K^{\mu\nu}$ (cf. Eq.~\eqref{eq: K kernel}), taking the divergence can introduce a delta function, resulting in a boundary term,
\begin{align}
\label{eq: gapless anomaly}
\begin{split}
    \partial_\mu K^{\mu\nu}&=-\,\delta(y)\,\Pi^{y\nu}(r,r')=\frac{i}{8\pi}\, \varepsilon^{y\nu a}\, \partial_a\, \delta^{(2)}(\rho-\rho')\, \delta(y)\, \delta_-(y'),
    \end{split}
\end{align}
where we used Eq.~\eqref{eq: boundary polarization tensor}. Thus, Eq.~\eqref{eq: matter resp gauge trans} is equivalent to
\begin{equation}
    \Delta S_{\mathrm{matter}}=-\, \frac{i}{8\pi}\int_{y\, =\, 0}d^2 \rho \, \xi \, F_{\tau x},
\end{equation}
which exactly cancels Eq.~\eqref{eq: level 1/2 CS gauge trans}. Thus, the full response $S_{\mathrm{resp}}[A_\mu]$ in Eq.~\eqref{eq: response} is gauge invariant.

We may equivalently demonstrate anomaly cancellation by examining the current generated in response to the background field $A_\mu$. By carefully calculating the variation of the action, we find that the current associated with the Chern-Simons term is
\begin{equation}
    \langle J^\mu(r) \rangle_{\mathrm{CS}}=\frac{\delta S_\mathrm{CS}}{\delta A_\mu(r)}=\frac{i}{4\pi}\, \Theta(-y)\, \varepsilon^{\mu\nu\lambda}\, \partial_\nu A_\lambda+\frac{i}{8\pi}\, \delta(y)\, \varepsilon^{y\mu \nu}\, A_\nu.
\end{equation}
The current associated with the gapless matter is
\begin{equation}
    \langle J^\mu(r) \rangle_{\mathrm{matter}}=\frac{\delta S_\mathrm{matter}}{\delta A_\mu(r)}=-\int_{y'\, <\, 0} d^3 r' \,K^{\mu\nu}(r,r')\,A_\nu(r').
\end{equation}
These currents are not conserved separately since
\begin{equation}
     \partial_\mu\langle J^\mu(r) \rangle_{\mathrm{CS}}=-\, \frac{i}{8\pi}\, \delta(y)\, F_{\tau x}, \qquad \partial_\mu\langle J^\mu(r) \rangle_{\mathrm{matter}}=\frac{i}{8\pi}\, \delta(y)\, F_{\tau x},
\end{equation}
where we used Eq.~\eqref{eq: gapless anomaly}. However, we see that the full current,
\begin{equation}
    \partial_\mu\langle J^\mu(r) \rangle_A=0,\qquad \langle J^\mu(r) \rangle_A=\langle J^\mu(r) \rangle_{\mathrm{matter}}+\langle J^\mu(r) \rangle_{\mathrm{CS}}
\end{equation}
is indeed conserved, thus explicitly demonstrating the anomaly cancellation.

This mechanism of anomaly cancellation should be contrasted with anomaly inflow in the gapped Chern insulator phase. In the gapped phase, the bulk carries a Chern-Simons response of level 1, which has an anomaly that is canceled by the chiral mode localized at the edge of the system. In contrast, at the quantum critical point between the Chern insulator and the trivial insulator, the chiral mode becomes delocalized. In this case, as we have shown above, the anomaly of the level 1/2 Chern-Simons response above is exactly canceled by the delocalized chiral mode in the critical bulk.

Finally, we comment on the relation of Eq.~\eqref{eq: response} to the APS eta invariant mentioned in Section~\ref{sec: physical setup}. The time-reversal odd terms of Eq.~\eqref{eq: response}, which include both the Chern-Simons terms and the non-local time-reversal odd terms in $\Pi^{\mu\nu}$, may be regarded as an approximation of the eta invariant, which is gauge invariant, to quadratic order in $A_\mu$. Indeed, we have expanded the effective action generated by integrating out the massless Dirac fermion in a half-space with a flat metric and weak background field $A_\mu$, which is sufficient for studying the perturbative anomaly as we did above, but more refined questions about global anomalies would require the full eta invariant.

\section{Thermal response}

\label{sec: thermal response}

Coupling the system to a gravitational background provides a formal way to describe thermal response~\cite{Luttinger1964,Volovik1990,Read2000,Ryu2012}. At a Chern insulator phase with Chern number $C$, the electrical Hall conductivity is accompanied by a universal low-temperature thermal Hall conductivity,
\begin{equation}
	\kappa_{xy}=C\,\frac{\pi k_B^2}{6\hbar}\,T.
\end{equation}
As reviewed in Appendix~\ref{sec: gravity review}, this thermal response is associated with a gravitational Chern-Simons term of level $k_g=2\,C$. At the transition between phases with Chern numbers $C=0$ and $C=1$, the effective Chern number is $C=1/2$, so this quantum critical point has a thermal Hall conductivity,
\begin{equation}
\label{eq: quantum critical thermal hall}
    	\kappa_{xy}=\frac{1}{2}\,\frac{\pi k_B^2}{6\hbar}\,T.
\end{equation}
The gravitational Chern-Simons term is anomalous in the presence of a boundary, and at the quantum critical point, while there are no chiral modes localized at the edge, this anomaly is canceled by the delocalized chiral mode near the edge. Following methods similar to those in Section~\ref{sec: currents and anomalies}, we explicitly demonstrate this anomaly matching mechanism below.

\subsection{Energy-momentum tensor correlation functions}

\label{sec: em tensor correlation functions}

We previously analyzed the two-point function of the $U(1)$ current and found a nonlocal parity-odd term associated with the $U(1)$ anomaly of the delocalized chiral mode. The analogue of the $U(1)$ current for the gravitational response is the energy-momentum tensor $T_{\mu\nu}$, which couples to a background metric.\footnote{For a review, see Refs.~\cite{DiFrancesco1997,Fradkin-2021}.} We therefore first analyze general constraints on the two-point function of the energy-momentum tensor for a (2+1)d CFT in a half-space that breaks time-reversal symmetry. For the parity-even terms~\cite{McAvity1993,McAvity1995,Herzog2017}, it is useful to define the tensors,
\begin{align}
\alpha_{\mu\nu}&=X_\mu \, X_\nu-\frac{1}{3}\, \delta_{\mu\nu}, \qquad \alpha'_{\mu\nu}=X_\mu'\, X_\nu'-\frac{1}{3}\, \delta_{\mu\nu},\\
\begin{split}
\beta_{\mu\nu,\lambda\sigma}&=X_\mu\, X_\sigma'\, I_{\nu\lambda}(r_-)+X_\nu\, X_\sigma'\, I_{\mu\lambda}(r_-)+X_\mu \, X_\lambda'\, I_{\nu\sigma}(r_-)+X_\nu\, X_\lambda'\, I_{\mu\sigma}(r_-)\\
&\hspace{5mm}-\frac{4}{3}\, \delta_{\lambda\sigma}\, X_\mu\, X_\nu-\frac{4}{3}\, \delta_{\mu\nu}\, X_\lambda'\, X_\sigma '+\frac{4}{9}\, \delta_{\mu\nu}\,\delta_{\lambda\sigma},
\end{split}\\
\mathcal{I}_{\mu\nu,\lambda\sigma}(r_-)&=\frac{1}{2}\left[I_{\mu\sigma}(r_-)\, I_{\nu\lambda}(r_-)+I_{\mu\lambda}(r_-)\, I_{\nu\sigma}(r_-)\right]-\frac{1}{3}\, \delta_{\mu\nu}\, \delta_{\lambda\sigma},
\end{align}
where $I_{\mu\nu}(r_-)$, $X_\mu$, and $X'_\mu$ are defined in Eqs.~\eqref{eq: I tensor} and \eqref{eq: X vectors}. For the parity-odd terms, it is useful to define
\begin{align}
    \mathcal{P}^{\mu\nu}&=\varepsilon^{\mu\lambda\sigma}\, X_\lambda\, {I_\sigma}^\nu(r_-),\\
\mathcal{Q}_1^{\mu\nu,\lambda\sigma}&=\mathcal{P}^{\mu\lambda}\, I^{\nu\sigma}(r_-)+\mathcal{P}^{\mu\sigma}\, I^{\nu\lambda}(r_-)+\mathcal{P}^{\nu\lambda}\, I^{\mu\sigma}(r_-)+\mathcal{P}^{\nu\sigma}\, I^{\mu\lambda}(r_-),\\
\mathcal{Q}_2^{\mu\nu,\lambda\sigma}&=\mathcal{P}^{\mu\lambda}\, X^\nu\, X'^\sigma+\mathcal{P}^{\mu\sigma}\,X^\nu\, X'^\lambda+\mathcal{P}^{\nu\lambda}\, X^\mu\, X'^\sigma+\mathcal{P}^{\nu\sigma}\, X^\mu\, X'^\lambda.
\end{align}
The most general two-point function for the energy-momentum tensor of a (2+1)d CFT in a half-space that breaks time-reversal symmetry is
\begin{align}
\label{eq: generic em tensor}
	\begin{split}
\langle T_{\mu\nu}(r) \,T_{\lambda\sigma}(r') \rangle&=\frac{1}{|r_-|^6}\Big[\,\alpha_{\mu\nu}\, \alpha_{\lambda\sigma}'\, A(v)+\beta_{\mu\nu,\lambda\sigma}\, B(v)+\mathcal{I}_{\mu\nu,\lambda\sigma}(r_-)\, C(v)\\
&\hspace{17mm}+i\hspace{.25mm} \mathcal{Q}_1^{\mu\nu,\lambda\sigma}\, Q_1(v)+i\hspace{.25mm}\mathcal{Q}_2^{\mu\nu,\lambda\sigma}\, Q_2(v)\,\Big]\\
&\hspace{5mm}-\frac{i\hspace{.25mm} k_g}{192\pi}\left\lbrace \left[\varepsilon_{\mu\lambda\rho} \,\partial^\rho\,(\partial_\nu \,\partial_\sigma- \delta_{\nu\sigma}\,\partial^2)+(\mu \leftrightarrow\nu)\right]+(\lambda\leftrightarrow\sigma)\right\rbrace \delta^{(3)}(r_-) .
\end{split}
\end{align}
The terms on the first line are even under parity and time-reversal and are well-studied in the boundary criticality literature~\cite{McAvity1995,McAvity1993,Herzog2017}. The third line is related to the gravitational Chern-Simons term. The terms on the second line are odd under parity and time-reversal and have not been analyzed previously.

Conservation of the energy-momentum tensor,
\begin{equation}
    \partial_\mu \langle T^{\mu\nu}(r)T^{\lambda\sigma}(r')\rangle=0,
\end{equation}
for $y<0$ and $y'<0$, imposes that~\cite{McAvity1995,McAvity1993,Herzog2017}
\begin{align}
\begin{split}
	\left(v\, \frac{d}{dv}-3\right)\left[C(v)+2\,B(v)\right]&=-\,\frac{2}{3}\,\left[A(v)+4\, B(v)\right]-3\,C(v),\\
		\left(v\, \frac{d}{dv}-3\right)\left[A(v)+B(v)\right]&=A(v)-5\,B(v).
            \end{split}
\end{align}
Similarly, for the parity-odd terms, we have the constraint,
\begin{equation}
            	\left(v\, \frac{d}{dv}-3\right)\left[Q_1(v)+Q_2(v)\right]=-\,4\,Q_1(v).
\end{equation}
Since $Q_1$ and $Q_2$ are not independent, they may be expressed in terms of a single function $D(v)=[Q_1(v)+Q_2(v)]/4$ so that
\begin{equation}
	Q_1(v)=3\,D(v)-v\, D'(v),\qquad Q_2(v)=D(v)+v\, D'(v).
    \end{equation}
In the deep bulk limit, where $y\to-\infty$ and $y'\to-\infty$ with $y-y'$ fixed, we have
\begin{equation}
    A(0)=B(0)=Q_1(0)=Q_2(0)=D(0)=0,
\end{equation}
while $C(0)$ and $k_g$ are determined by the bulk CFT.

To analyze the gravitational anomaly, an important limit is
\begin{equation}
\label{eq: T corr bdry}
    \langle T_{ya}(\tau,x,y\to 0^-) \, T_{\lambda\sigma}(r')\rangle,
\end{equation}
where $a\in\lbrace \tau,x \rbrace$. Naively, \eqref{eq: T corr bdry} vanishes if
\begin{equation}
    2\,B(1)+C(1)=D(1)=0,
\end{equation}
but by methods similar to those of Appendix~\ref{sec: delocalized anomaly}, we find that the distributional limit contains a universal parity-odd contact term,
\begin{align}
\begin{split}
    &\langle T_{ya}(\tau,x,y\to 0^-) \, T_{\lambda\sigma}(r')\rangle\\
    &=\frac{i\hspace{.25mm}(k_D- k_g)}{384\pi}\left\lbrace \left[\varepsilon_{y\lambda b} \,\tilde{\partial}^b\,(\tilde{\partial}_a \,\tilde{\partial}_\sigma- \delta_{a\sigma}\,\tilde{\partial}^2)+(y \leftrightarrow a)\right]+(\lambda\leftrightarrow\sigma)\right\rbrace \delta^{(2)}(\rho-\rho')\,\delta_-(y'),
    \end{split}
\end{align}
where $\tilde{\partial}_\mu=(\partial_a,-\partial_{y'})$, $\tilde{\partial}^2=\partial_a^2+\partial_{y'}^2$, and we use indices $a,b\in\lbrace \tau, x\rbrace$. The anomaly coefficient $k_D$ is
\begin{equation}
\label{eq: grav anomaly coefficient}
    k_D=8\pi^2\int_{0}^{1}\frac{dv}{v^6}\left[\frac{4\,v^3}{1-v^2}-3\,(1-v^2)\,[(1+v^2)\,\operatorname{arctanh}(v)-v]\right]D(v),
\end{equation}
where we assume that $D(0)=D(1)=0$.

For the Chern insulator transition, the energy-momentum tensor is
\begin{equation}
\label{eq: energy-momentum tensor}
    T_{\mu\nu}=\frac{1}{4}\,\bar{\psi}\left(\Gamma_\mu \,\overleftrightarrow{\partial_\nu}+\Gamma_\nu \,\overleftrightarrow{\partial_\mu}\right)\psi.
\end{equation}
Applying Wick's theorem, the two-point function of the energy-momentum tensor for $y<0$ and $y'<0$ is
\begin{equation}
\label{eq: dirac fermion em tensor}
	\langle T^{\mu\nu}(r)\, T^{\lambda\sigma}(r')\rangle=\frac{1}{16}
	\left[\mathcal{C}^{\mu\nu,\lambda\sigma}(r,r')
	+\mathcal{C}^{\mu\nu,\sigma\lambda}(r,r')+\mathcal{C}^{\nu\mu,\lambda\sigma}(r,r')
	+\mathcal{C}^{\nu\mu,\sigma\lambda}(r,r')
	\right],
\end{equation}
where we define
\begin{equation}
\label{eq: C tensor}
	\mathcal{C}^{\mu\nu,\lambda\sigma}(r,r')
	=
	-\lim_{\substack{u,w\to r\\ u',w'\to r'}}(\partial_u^\nu-\partial_w^\nu)\,(\partial_{u'}^\sigma-\partial_{w'}^\sigma)
	\operatorname{Tr}\left[\Gamma^\mu\,
	 S_\pm(w,u')\,
	\Gamma^\lambda\,
	S_\pm(w',u)\right],
\end{equation}
and we use Eq.~\eqref{eq: fermion correlations both bc} for the fermion propagator $S_\pm$. Calculating Eq.~\eqref{eq: dirac fermion em tensor} using Eqs.~\eqref{eq: C tensor} and \eqref{eq: fermion correlations both bc} yields a two-point function for the energy-momentum tensor of the form in Eq.~\eqref{eq: generic em tensor} with
\begin{gather}
\label{eq: CI em tensor funcs}
   A(v)=\frac{3}{4\pi^2}\, v^6, \qquad B(v)=-\, \frac{3}{16\pi^2}\, v^6,\qquad C(v)=\frac{3}{16\pi^2}\,(1+v^6),\\
   D(v)=\pm\,\frac{3}{64\pi^2}\,v^2\,(1-v^2),\qquad k_g=0, \nonumber
\end{gather}
which indeed satisfy $A(0)=B(0)=D(0)=0$ and $2\, B(1)+C(1)=D(1)=0$. Applying Eq.~\eqref{eq: grav anomaly coefficient} to $D(v)$ in Eq.~\eqref{eq: CI em tensor funcs} yields an anomaly coefficient of $k_D=\pm1$. Hence, for the right-moving boundary condition in Eq.~\eqref{eq: fermion bc}, we have $k_D=1$. The nonlocal parity-odd part of the energy-momentum tensor two-point function carries the gravitational anomaly coefficient of the delocalized chiral mode. To establish anomaly inflow, this contribution must be combined with the local gravitational response of the heavy fermion doubler. In the following subsection, we show explicitly that the anomalies of these two contributions cancel.

\subsection{Gravitational anomaly}

\label{sec: gravitational anomaly}

We now demonstrate gravitational anomaly inflow by combining the response of the gapless Dirac fermion with the gravitational Chern-Simons response of the heavy fermion doubler. We couple the system to a metric that weakly deviates from flat spacetime,
\begin{equation}
g_{\mu\nu}=\delta_{\mu\nu}+h_{\mu\nu},
\end{equation}
where $|h_{\mu\nu}|\ll 1$. Our conventions for curved spacetime and the coupling of fermions to a background metric are reviewed in Appendix~\ref{sec: gravity review}.

As further reviewed in Appendix~\ref{sec: gravity review}, to quadratic order in $h_{\mu\nu}$, the contribution of the heavy fermion doubler to the gravitational response is the approximation to the level 1 gravitational Chern-Simons term,
\begin{equation}
	\label{eq: grav by parts}
	 S_{\mathrm{gCS}}[h_{\mu\nu}]=\frac{i}{384 \pi}\int_{y\,<\,0} d^3 r \, \varepsilon^{\mu\nu\lambda}\, h_{\mu\sigma}\, \partial_\nu \left(\partial^2\, {h_\lambda}^\sigma-\partial^\sigma \,\partial_\rho \,{h_\lambda}^\rho\right).
\end{equation}
The contribution from the gapless Dirac fermion is
\begin{equation}
    S_{\mathrm{matter}}[h_{\mu\nu}]=-\,\frac{1}{8}\int_{y\, < \, 0} d^3 r \int_{y'\, < \, 0} d^3 r'\, h_{\mu\nu}(r)\,\langle T^{\mu\nu}(r)\, T^{\lambda\sigma}(r') \rangle  \,h_{\lambda\sigma}(r'),
\end{equation}
where $\langle T^{\mu\nu}(r)\, T^{\lambda\sigma}(r')\rangle$ is given in Eq.~\eqref{eq: dirac fermion em tensor}. An infinitesimal diffeomorphism acts in linearized gravity as the gauge transformation,
\begin{equation}
\label{eq: inf diffeomorphism}
    h_{\mu\nu}\to h_{\mu\nu}+\partial_\mu \xi_\nu+\partial_\nu \xi_\mu,
\end{equation}
where we take $\xi_y|_{y=0}=0$ to preserve the boundary. Under the gauge transformation, the gravitational Chern-Simons term in Eq.~\eqref{eq: grav by parts} changes by the boundary term,
\begin{equation}
\label{eq: gCS anomaly}
	 \Delta S_{\mathrm{gCS}}=-\,\frac{i}{96 \pi}\int_{y\,=\,0} d^2 \rho \, \xi_a\, C^{ya}, \qquad C^{\mu\nu}=\varepsilon^{\mu\lambda\sigma}\,\partial_\lambda \left({R_\sigma}^\nu-\frac{1}{4}\,R\,{\delta_\sigma}^\nu\right),
\end{equation}
where $R_{\mu\nu}$ and $R$ are the Ricci curvature tensor and Ricci scalar in linearized gravity, respectively, given by
\begin{equation}
    	R_{\mu\nu}=\frac{1}{2}\left(\partial_\sigma \,\partial_\nu \,{h^\sigma}_\mu+\partial_\sigma \, \partial_\mu \, {h^\sigma}_\nu-\partial_\mu \,\partial_\nu \,{h_\sigma}^\sigma-\partial^2 \,h_{\mu\nu}\right), \qquad R={R_\mu}^\mu.
\end{equation}
Meanwhile, by analogous reasoning to that in Section~\ref{sec: anomaly inflow}, the response due to the gapless matter changes under Eq.~\eqref{eq: inf diffeomorphism} by
\begin{align}
\label{eq: gapless matter grav anomaly}
	\Delta S_{\mathrm{matter}}=\frac{i}{96\pi}\,(k_D-k_g)\int_{y\,=\,0} d^2 \rho \,\xi_a\, C^{ya}.
\end{align}
We have $k_g=0$, and if we take the right-moving boundary condition, Eq.~\eqref{eq: fermion bc}, then we find $k_D=1$. Then the anomaly of the gravitational Chern-Simons term, Eq.~\eqref{eq: gCS anomaly}, is exactly canceled by Eq.~\eqref{eq: gapless matter grav anomaly}. Thus, we have explicitly demonstrated that the gravitational anomaly associated with the thermal Hall response in Eq.~\eqref{eq: quantum critical thermal hall} is matched by the delocalized chiral mode at the critical point, even though no chiral mode remains exponentially localized at the edge.

Finally, we comment that our analysis in this section is easily generalizable to the transition between a trivial superconductor with chiral central charge $c_-=0$ and a $p_x+i p_y$ superconductor with chiral central charge $c_-=1/2$~\cite{Read2000}. This transition is described by a single Majorana fermion whose mass changes sign at the transition, and the thermal Hall conductivity deep in the bulk is half of Eq.~\eqref{eq: quantum critical thermal hall}. The two-point function for the energy-momentum tensor is of the form in Eq.~\eqref{eq: generic em tensor} but with functions that are half of those in Eqs.~\eqref{eq: CI em tensor funcs}, and accordingly, we find $k_D=1/2$. Hence, at this transition there is also a chiral mode that delocalizes from the edge and cancels the anomaly of the gravitational Chern-Simons term.

More broadly, as emphasized in Section~\ref{sec: em tensor correlation functions}, we may define $k_D$ for an arbitrary (2+1)d CFT in a half-space. Since this coefficient quantifies the gravitational anomaly of the chiral mode that delocalizes from the edge, it is natural to regard $k_D/2$ as an analog of the chiral central charge for the delocalized chiral mode. However, one must keep in mind that this anomaly is not a property of the (1+1)d edge theory but rather of the bulk CFT and the boundary conditions.

\section{Higher Chern number transitions}

\label{sec: generalized chern transitions}

Thus far, in this work we have only considered a transition between a Chern insulator with $C=1$ and a trivial insulator. In this section, we generalize our results to other kinds of Chern insulator transitions. Edge modes of transitions of this type were previously discussed in Ref.~\cite{Verresen2020}, and our results below are in agreement. Here, we will explain how to obtain the edge physics in our construction with fermion mass domain walls and how to understand anomaly matching for these critical points.

We consider a transition between Chern insulators with Chern numbers $C=k$ and $C=k+\Delta k$, where $k\geq 0$ and $\Delta k>0$ are integers. The bulk transition is described by $\Delta k$ Dirac fermions whose masses change sign at the transition. To have a direct Chern number changing transition with $\Delta k>1$, it is necessary to impose an additional symmetry (e.g., a lattice symmetry). Otherwise, we generically expect that $\Delta k=1$, but here, we will keep $\Delta k\geq 1$ generic. To model the boundary physics of this critical point, we take $k+\Delta k$ Dirac fermions with the action,
\begin{equation}
\label{eq: higher chern transition}
    S=\int d^3 r\, \sum_{j=1}^{\Delta k}\bar{\psi}_j\,[i\hspace{.25mm}\slashed{\partial}-M(y)]\,\psi_j+\int d^3 r\, \sum_{J=1}^{k}\overline{\Psi}_J\,[i\hspace{.25mm}\slashed{\partial}-M_0(y)]\,\Psi_J,
\end{equation}
where the mass profiles are given by
\begin{equation}
\label{eq: higher chern masses}
    M(y)=-\, m_0\, \Theta(y), \qquad M_0(y)=-\, m_0\, \Theta(y)+m\, \Theta(-y),
\end{equation}
where $0<m<m_0$. Hence, we have $k$ fermions $\Psi_J$ that have mass domain walls across which their Dirac masses change sign, and there are $\Delta k$ fermions $\psi_j$ with a step function mass. Although we take the fermions to have the same mass $-m_0$ for $y>0$ for simplicity, this symmetry is not necessary for any essential physics. Similarly, it is also not essential to assume that the $k$ fermions have the same mass $m$ for $y<0$.

If the $\Delta k$ fermions are instead given a nonzero mass for $y<0$, then depending on the sign of this mass, there will either be $k$ localized chiral edge modes or $k+\Delta k$ chiral edge modes, signaling the respective Chern insulator phases. Thus, this fermion mass domain wall model indeed captures the physics of this transition. There are also $k+\Delta k $ heavy fermion doublers, which are included to properly quantize the Chern number in each phase.

Repeating the analysis of Section~\ref{sec: two-point function} for the model in Eq.~\eqref{eq: higher chern transition}, in the $m_0,m\to\infty$ limit we find that at the critical point there are $k$ localized chiral edge modes with edge correlation functions,
\begin{equation}
    \langle \Psi_I(r) \, \overline{\Psi}_J(0)\rangle\, |_{y=0}\sim \delta_{IJ}\,\frac{\Gamma^+}{x^+},
\end{equation}
which are the same as the edge modes in a gapped Chern insulator phase. There are also $\Delta k$ delocalized modes at the transition with edge correlation functions,
\begin{equation}
    \langle \psi_i(r) \, \bar{\psi}_{j}(0)\rangle\, |_{y=0}\sim \delta_{ij}\,\frac{\Gamma^+}{|\rho| \, x^+}.
\end{equation}
Each of these modes represents a chiral fermion with an anomalous dimension as in Eq.~\eqref{eq: boundary propagator}.

As in Section~\ref{sec: anomaly inflow}, we must study the response to a background electromagnetic field $A_\mu$. We again take $m,m_0\to\infty$ and include the terms from the heavy fermion doublers so that the $y>0$ region is a trivial insulator. The resulting bulk response to the background electromagnetic field $A_\mu$ is
\begin{align}
\begin{split}
    S_\mathrm{bulk}[A_\mu]&=-\frac{\Delta k}{2}\int_{\mathbb{R}^3} d^3 r \int_{\mathbb{R}^3} d^3 r' \,A_\mu(r)\, K^{\mu\nu}(r,r')\, A_\nu(r')+\frac{i(k+\Delta k/2)}{4\pi}\int_{y\, <\, 0} A\, dA.
\end{split}
\end{align}
This bulk effective action is not gauge invariant on its own, but coupling to the $k$ chiral fermions localized at the edge leads to a gauge invariant response. 

While we have modeled this Chern insulator transition using the mass profiles in Eq.~\eqref{eq: higher chern masses}, other kinds of boundary conditions are possible. As discussed in Ref.~\cite{Verresen2020}, there is a notion of bulk-boundary correspondence that allows one to compute a topological invariant to check what sets of chiral modes are compatible with the bulk critical point. The effective Chern number of the bulk critical point is related to the number of chiral modes by
\begin{equation}
\label{eq: top invariant}
    C=N^R_\mathrm{localized}-N^L_\mathrm{localized}+\frac{1}{2}\left( N^R_\mathrm{delocalized}-N^L_\mathrm{delocalized}\right),
\end{equation}
where $N^{R/L}_\mathrm{localized}$ denotes the number of right/left-moving chiral modes localized at the edge and $N^{R/L}_\mathrm{delocalized}$ denotes the number of right/left-moving chiral modes delocalized into the bulk. 

Hence, Eq.~\eqref{eq: top invariant} predicts that it is possible to have a different number of localized and delocalized modes as long as the effective Chern number remains the same. For example, at a transition where the Chern number changes from $C=0$ to $C=1$, there can be one delocalized right-moving mode, which is the case we have treated for most of this work. Alternatively, we could have one right-moving mode localized at the edge and one delocalized left-moving mode since the net contribution is $C=1-1/2=1/2$. Within our framework, the generalized notion of bulk-edge correspondence for critical points may be understood directly from gauge invariance. Indeed, as long as two different combinations of localized and delocalized chiral modes lead to a gauge invariant electromagnetic response (and gravitational response), they may possibly be realized at the critical point. Just as in gapped phases, anomaly matching does not uniquely fix a boundary state but places strong restrictions on what boundary physics is possible. Identifying the boundary state realized in a particular microscopic model requires additional, non-universal information.

Finally, we comment on the robustness of the edge modes. Suppose the system is infinite (or periodic) in the $x$-direction but is of finite length $L$ in the $y$-direction with open boundary conditions. As observed in Ref.~\cite{Verresen2020}, while the chiral modes localized at the edge are protected, the delocalized chiral modes are less robust because currents of opposite chiralities associated with the two boundaries can mix with one another. To observe the effects of the delocalized chiral current we have explored in this work, $L$ must be much greater than the correlation length $\xi$. Thus, there is an important order of limits: We must take the thermodynamic limit, $L\to\infty$, and then tune to criticality, $\xi\to\infty$. From the theoretical standpoint, this limit corresponds to working in a semi-infinite half-space as we have done throughout this work. This subtlety is not unique to the present problem but arises in the theory of boundary criticality more generally~\cite{Diehl1997}.

\section{(3+1)d topological insulator boundary criticality}

\label{sec: TI boundary criticality}

Having solved for the boundary physics of the Chern insulator transition, we can easily generalize our methods to solve for the boundary physics of a transition between a (3+1)d time-reversal invariant (strong) topological insulator (TI)~\cite{Fu2007,Moore2007,Roy2009} and a trivial insulator. In the topological insulator phase, the bulk is gapped, but as long as time-reversal symmetry and $U(1)$ charge conservation are preserved, the surface state has a single Dirac cone that cannot be gapped (at least without interactions). Since this free fermion surface state may be understood by considering a domain wall across which the mass of a (3+1)d Dirac fermion changes sign~\cite{Jackiw1976,Boyanovsky1987,Kaplan1992,Qi-2008,Mulligan2013}, we may analyze the boundary of a transition between a (3+1)d TI and a trivial insulator in a similar fashion to our analysis of the Chern insulator transition.

We take the domain wall to be perpendicular to the $z$-direction and consider a Dirac mass profile given by
\begin{equation}
\label{eq: (3+1)d mass domain wall}
    M(z)=\begin{cases}
        m & z<0,\\
        -\, m_0 & z>0,
    \end{cases}
\end{equation}
where $0<m \ll m_0$. In Appendix~\ref{sec: (3+1)d mass domain wall}, we solve for the fermion propagator in Euclidean signature for $m_0\to\infty$ and $m\geq 0$. In the $m_0\to\infty$ limit, the four-component Dirac spinor obeys the boundary condition,\footnote{Unlike in (2+1)d, the boundary conditions $-\Gamma^z\psi(\tau,x,y,z=0)=\pm \, \psi(\tau,x,y,z=0)$ are not the only ones consistent with the residual conformal symmetry parallel to the boundary. The most general boundary condition consistent with this symmetry is $U(\theta)\,\psi(\tau,x,y,z=0)=\psi(\tau,x,y,z=0)$ where $U(\theta)=\Gamma^z\, e^{i\theta \Gamma^5}$, $0\leq \theta<2\pi$, and $\Gamma^5=\Gamma^\tau \Gamma^x\Gamma^y\Gamma^z$. Further restricting to boundary conditions consistent with time-reversal symmetry requires $\theta=0,\pi$. See Refs.~\cite{Kurkov2017,Kurkov2018} for an analysis of anomalies for alternate boundary conditions with $\theta=\pm \pi/2$, which is consistent with the residual conformal symmetry but violates time-reversal symmetry.}
\begin{equation}
    -\Gamma^z\, \psi(\tau,x,y,z=0)=\psi(\tau,x,y,z=0).
\end{equation}
For $m=0$, the method of images works, and the solution for the fermion propagator (in Euclidean signature) for $z\leq 0$ and $z'\leq 0$ is
\begin{equation}
\label{eq: TI propagator}
    \langle \psi(r)\, \bar{\psi}(r')\rangle=S_0^{\mathrm{4d}}(r_-)+ S_0^\mathrm{4d}(r_+)\,\Gamma^z,
\end{equation}
where $(r_\pm)_\mu=(\tau-\tau',x-x',y-y',z\pm z')$ and the propagator for a massless Dirac fermion in 4d Euclidean spacetime with no boundary is
\begin{equation}
	S_0^\mathrm{4d}(r)=\frac{\Gamma^\mu\, r_\mu}{2\pi^2 |r|^4}.
\end{equation}
Our convention for Dirac matrices in Euclidean signature is
\begin{equation}
	\Gamma^a=\begin{pmatrix}
		0&-\,i \hspace{.25mm}\sigma^a\\i\hspace{.25mm} \sigma^a &0
	\end{pmatrix}, \qquad \Gamma^z=\begin{pmatrix}
	\mathbb{I}_2&0\\0&-\,\mathbb{I}_2
	\end{pmatrix},
\end{equation}
where the indices $a\in\left\lbrace \tau,x,y\right\rbrace $ correspond to Pauli matrices $\sigma^1$, $\sigma^2$, $\sigma^3$ respectively.

Taking the boundary limit of the propagator in Eq.~\eqref{eq: TI propagator} gives
\begin{equation}
\label{eq: TI bdry propagator}
	\langle \psi(r)\, \bar{\psi}(0)\rangle|_{z=0}=\frac{\Gamma^a \, (r_\parallel)_a}{\pi^2 |r_\parallel|^4} \frac{\mathbb{I}_4+\Gamma^z}{2}=\frac{i}{\pi^2 |r_\parallel|^4}\begin{pmatrix}
		0&0\\ \sigma^a \, (r_\parallel)_a&0
	\end{pmatrix},
\end{equation}
where $(r_\parallel)_a=(\tau,x,y)$ and $|r_\parallel|=\sqrt{\tau^2+x^2+y^2}$. As in Section~\ref{sec: chern insulator near criticality}, if $m$ is positive but small, then the boundary two-point function is
\begin{equation}
\langle \psi(r)\, \bar{\psi}(0)\rangle|_{z=0}=\frac{\Gamma^a\, (r_\parallel)_a}{\pi^2 |r_\parallel|^4}\left(1+\frac{\pi \,m\,|r_\parallel|}{4}\right)\frac{\mathbb{I}_4+\Gamma^z}{2}.
\end{equation}
For $m>0$, the second term dominates at low energies. This term represents the propagator for the (2+1)d Dirac cone localized at the surface of the gapped TI phase. For $m=0$, the fermion propagator at the boundary reduces to Eq.~\eqref{eq: TI bdry propagator}, which resembles a (2+1)d massless Dirac fermion that has acquired an anomalous dimension so that its scaling dimension matches that of the (3+1)d bulk Dirac field. These results closely parallel those of Section~\ref{sec: chern insulator near criticality}. Unlike the Chern insulator, the (3+1)d topological insulator can alternatively host a gapped topologically ordered surface state, known as the T-Pfaffian~\cite{Chen2014a}, in the presence of interactions. We leave a study of the behavior of the T-Pfaffian surface state as the bulk TI approaches criticality to future work.

\section{Discussion}

\label{sec: discussion}

In this work, we analyzed the boundary physics of a Chern insulator transition using a Dirac mass domain wall construction. Because of the parity anomaly, the bulk critical theory carries a Hall conductivity $\sigma_{xy}=e^2/2h$, and this half-integer Hall response must have physical consequences when the system has an edge. We demonstrated that these consequences are visible directly in boundary correlation functions and in anomaly inflow.

The simplest manifestation of this chirality appears in the boundary fermion correlation function. At the critical point, the boundary propagator has the same chiral matrix structure as the propagator of a (1+1)d chiral fermion, but it is dressed by the gapless bulk so that it has the scaling dimension of the bulk Dirac fermion. This observable provides a concrete physical consequence of the bulk half-integer Hall conductivity.

We also showed how this delocalized mode emerges from the gapped Chern insulator phase. In the gapped phase, when the bulk has Dirac mass $m$, the usual chiral edge mode is exponentially localized near the interface with a length scale controlled by $1/m$. As the bulk approaches criticality, this length scale diverges, and the chiral mode leaks into the bulk. At the quantum critical point, the exponential profile is replaced by a power law structure. 

The current correlation functions give a complementary characterization of the same physics. The boundary current correlator is purely chiral, but deep in the bulk, the currents have no net chirality. Another central result of this work is the explicit demonstration of anomaly inflow at the critical point. In a gapped Chern insulator, the anomaly of the bulk Chern-Simons response is canceled by a localized chiral edge mode. At the transition, we showed that the chiral mode delocalized in the bulk matches the anomaly associated with the half-integer Hall conductivity. More generally, we found that the delocalized chiral mode and its anomaly are encoded in parity-odd structures in current and energy-momentum tensor correlation functions that, to our knowledge, have not previously been systematically studied. The possibility of these structures may be deduced directly from the residual conformal symmetry of a CFT in the presence of a boundary. We used these symmetry-allowed terms to characterize the delocalized chiral mode and to calculate both its electromagnetic and gravitational anomaly coefficients for a general (2+1)d CFT in a half-space. We apply this analysis to the Chern insulator transition and generalize to the free Majorana CFT, which describes the transition between a trivial superconductor and a topological superconductor. Thus, even when there is no sharply separated boundary theory, anomaly matching may be realized by critical bulk degrees of freedom near the edge.

We further extended our analysis to other kinds of Chern insulator transitions, between phases with Chern numbers $C=k$ and $C=k+\Delta k$ for integers $k\geq 0$ and $\Delta k>0$. In that case, the boundary can contain both localized and delocalized chiral modes. The allowed combinations are constrained by the effective Chern number of the bulk critical point, but they are not uniquely fixed by anomaly matching alone. As in gapped phases, anomalies alone do not uniquely specify a boundary state but highly constrain which edge states are allowed.

Finally, we applied the same method to the transition between a (3+1)d topological insulator and a trivial insulator. The boundary fermion two-point function at this critical point resembles the Dirac cone surface state but with a modified power law so that the scaling dimension matches that of the bulk fermion. Unlike the Chern insulator transition, this bulk critical point respects time-reversal symmetry, and its boundary is not chiral. 

There are several natural directions for future work. One is to investigate boundary states of fractional quantum Hall transitions, which we will present in an upcoming work. Another direction is to incorporate the effects of disorder, which are important in realistic experimental settings. Finally, in the (3+1)d topological insulator problem, interactions allow gapped topologically ordered surface states such as the T-Pfaffian~\cite{Chen2014a}. Understanding how such a surface state evolves as the bulk approaches criticality remains an important open problem.

\textit{Dedication:} This paper is dedicated to the memory of our colleague and friend Ian Affleck who passed away nearly two years ago. Ian was a true giant of theoretical physics. His deep insights and ideas made an enormous and long lasting impact on our field. This paper is a small contribution in his honor in an area of physics he really cared about.

\textit{Note added:} After this work was posted, we became aware of an independent work by M. Zeng~\cite{Zeng2026} that computes an information theoretic quantity known as the modular commutator for Chern insulator quantum critical points, providing a complementary perspective on how to characterize the chirality of edge degrees of freedom at criticality.

\section*{Acknowledgments}
We thank Hao-Ran Cui, Hart Goldman, and Alex Thomson for useful discussions and collaboration on a related project. We also thank Barry Bradlyn, Junyi Cao, Stefan Divic, Thomas Faulkner, Yin-Chen He, Jainendra Jain, Jaewon Kim, Kyung-Su Kim, Abijith Krishnan, Ethan Lake, John McGreevy, Vadim Oganesyan, Amir Raz, Subir Sachdev, Thomas Scaffidi, Sounak Sinha, Michael Stone, Ruben Verresen, Pengjie Wang, and Deyi Zhuo for useful discussions. This work was supported in part by the National Science Foundation under grant DMR-2225920 at the University of Illinois.

\appendix

\section{Fermion two-point function for finite \texorpdfstring{$m_0$}{m0}}

\label{sec: finite m0 propagator}

In this appendix, we solve for a complete basis of solutions to the eigenvalue equation, Eq.~\eqref{eq: matrix eigenvalue eq}, for finite $m_0>0$. In parallel with Section~\ref{sec: method of images}, we subsequently use these solutions to construct the fermion propagator for finite $m_0$.

As discussed in Section~\ref{sec: step function mass}, the eigenvalues are parametrized by Eq.~\eqref{eq: dirac eigenvalues}, and there we determined solutions for $\phi_\lambda(y)$ with $0\leq q \leq m_0$. Hence, we must now consider scattering eigenfunction solutions to Eqs.~\eqref{eq: matrix eigenvalue eq} and \eqref{eq: tise} with $q>m_0$. There are two classes of solutions---those with incident waves from $y\to-\infty$ and those with incident waves from $y\to\infty$. For $y<0$, the solutions with plane waves incident from $y\to-\infty$ are
\begin{equation}
	\phi^L_{\sigma,q}(y)=\frac{1}{2 m_0\sqrt{\pi \omega_q(\omega_q+\sigma\, p_x)}}\begin{pmatrix}
		iq\left[m_0\, e^{iq y}-i\hspace{.25mm}(q-k_q)\,e^{-iqy}\right]\\
		(p_x+\sigma \, \omega_q)\left[m_0 \,e^{iq y}+i\hspace{.25mm}(q-k_q)\,e^{-iqy}\right]
	\end{pmatrix},
\end{equation}
where $\omega_q=\sqrt{p_x^2+q^2}$, as defined in Section~\ref{sec: step function mass}, and we further define $k_q=\sqrt{q^2-m_0^2}$. For $y>0$, this solution is
\begin{equation}
	\phi^L_{\sigma,q}(y)=\frac{1}{2 m_0\sqrt{\pi \omega_q(\omega_q+\sigma\, p_x)}}\,e^{ik_q y}\begin{pmatrix}
		i\hspace{.25mm}q\hspace{.25mm} (m_0-iq+ik_q)\\
		(p_x+\sigma \, \omega_q)(m_0+i\hspace{.25mm}q-i\hspace{.25mm}k_q)
	\end{pmatrix}.
\end{equation}
We denote the other class of solutions, with waves incident from $y\to\infty$, by $\phi_{\sigma,q}^R(y)$. For $y<0$, we have 
\begin{equation} \phi_{\sigma,q}^R(y)=\frac{1}{\sqrt{2\pi \omega_q(\omega_q+\sigma\, p_x)}}\frac{\left(-k_q^2+q\, k_q\right)^{1/2}}{m_0}\, e^{-iqy}\begin{pmatrix} q\\ i(p_x+\sigma \, \omega_q )\end{pmatrix}. 
\end{equation}
For $y>0$, the solution is
\begin{equation} \phi_{\sigma,q}^R(y)=\frac{1}{\sqrt{2\pi \omega_q(\omega_q+\sigma\, p_x)}}\frac{\left(-k_q^2+q\, k_q\right)^{1/2}}{m_0}\begin{pmatrix} q\left[\cos(k_q y)+\frac{m_0-iq}{k_q}\sin(k_q y)\right]\\i (p_x+\sigma \, \omega_q)\left[\cos(k_q y)-\frac{m_0+iq}{k_q}\sin(k_q y)\right] \end{pmatrix}. \end{equation}
The eigenfunctions are normalized so that
\begin{gather}
    \int_{-\infty}^\infty dy\left[\phi^L_{\sigma,q}(y)\right]^\dagger \phi^L_{\sigma',q'}(y)=\int_{-\infty}^\infty dy\left[\phi^R_{\sigma,q}(y)\right]^\dagger \phi^R_{\sigma',q'}(y)=\delta_{\sigma \sigma'}\,\delta(q-q'),\nonumber \\
    \int_{-\infty}^\infty dy\left[\phi^R_{\sigma,q}(y)\right]^\dagger \phi^L_{\sigma',q'}(y)=0.
\end{gather}
Together with the solutions for $0\leq q\leq m_0$ presented in Section~\ref{sec: step function mass}, these eigenfunctions of $\mathcal{D}$ form a complete basis.

Using these eigenfunctions, we may then determine the fermion two-point function from Eq.~\eqref{eq: propagator spectral}. After computing the integrals, we find that the fermion propagator in Euclidean spacetime for $y\leq 0$ and $y'\leq 0$ is
\begin{align}
	\begin{split}
	\langle \psi(r)\, \bar{\psi}(r')\rangle&=\frac{(r_-)_\mu \Gamma^\mu}{4\pi |r_-|^3}+\frac{(r_+)_\mu \Gamma^\mu\Gamma^y 3(y+y')-\mathbb{I}_2 |r_+|^2}{4\pi m_0 |r_+|^5}-\frac{(r_-)_a\Gamma^a \Gamma^y}{4\pi m_0 |\rho|} \partial_{|\rho|} \,\mathcal{I}-\frac{\mathbb{I}_2}{4\pi m_0} \partial_y\, \mathcal{I},
	\end{split}
\end{align}
where $\mu\in\lbrace\tau,x,y\rbrace$, $a\in\lbrace\tau,x\rbrace$, and the remaining integral is
\begin{align}
\mathcal{I}(|\rho|,  y+y',m_0)=\int_{0}^{\infty}dk\,\sqrt{k^2+m_0^2}\,e^{-k|y+y'|}J_0(|\rho| \hspace{.25mm} k),
\end{align}
and $J_0$ is a Bessel function of the first kind. For $y\geq 0$ and $y'\geq 0$, the fermion two-point function is
\begin{align}
	\begin{split}
	\langle \psi(r)\, \bar{\psi}(r')\rangle&=\frac{e^{-m_0|r_-|}}{4\pi |r_-|^3}(1+m_0 |r_-|)(r_-)_\mu \Gamma^\mu-\frac{e^{-m_0 |r_+|}}{4\pi |r_+|^3}(1+m_0|r_+|)(r_-)_a \Gamma^a\\
	&\hspace{5mm}-\frac{e^{-m_0 |r_+|}}{4\pi m_0 |r_+|^5}(r_+)_\mu \Gamma^\mu \Gamma^y(y+y')\left(3+3m_0|r_+|+m_0^2 |r_+|^2\right)\\
    &\hspace{5mm}+\mathbb{I}_2 \left[\frac{e^{-m_0 |r_+|}}{4\pi m_0 |r_+|^3}\left(1+m_0|r_+|+m_0^2 |r_+|^2\right)-\frac{m_0\, e^{-m_0 |r_-|}}{4\pi |r_-|}\right]\\
	&\hspace{5mm}+\frac{1}{4\pi m_0}\left[\mathbb{I}_2\left(\partial_{|\rho|}+\frac{1}{|\rho|}\right)+\frac{(r_-)_a \Gamma^a}{|\rho|}(\mathbb{I}_2\,m_0-\Gamma^y \partial_y)\right]\hspace{-1mm}\partial_{|\rho|}\partial_y \mathcal{J}(|r_+|,y+y',m_0),
\end{split}
\end{align}
where we define
\begin{equation}
    \mathcal{J}(|r_+|,y+y',m_0)=I_0\left(\frac{m_0}{2}  \left(y+y'-|r_+|\right)\right) K_0\left(\frac{m_0}{2}  \left(y+y'+|r_+|\right)\right),
\end{equation}
with $I_0$ and $K_0$ denoting modified Bessel functions. For finite $m_0>0$, there is no solution by the method of images, yet the fermion propagator is still tractable.

It is useful to check limiting cases. For $m_0\to\infty$, we recover Eq.~\eqref{eq: images solution} for $y\leq 0$, $y'\leq 0$, and the propagator vanishes for $y> 0$, $y'> 0$. In the $m_0\to 0$ limit, we find the free Dirac propagator,
\begin{equation}
\label{eq: massless bulk}
    S_0(r_-)=\frac{\Gamma^\mu (r_-)_\mu}{4\pi |r_-|^3},
\end{equation}
for both $y\geq0$, $y'\geq 0$ and $y\leq 0$, $y'\leq 0$. In the bulk limit on the massless side of the domain wall, we take $y\to -\infty$, $y'\to-\infty$ with $y-y'$ fixed and also recover Eq.~\eqref{eq: massless bulk}. On the other side, when we take $y\to\infty$ and $y'\to\infty$ with $y-y'$ fixed, we obtain
\begin{equation}
    S_\mathrm{massive}(r_-)=\frac{e^{-m_0 |r_-|}}{4\pi |r_-|^3}\left[\Gamma^\mu (r_-)_\mu\left(1+m_0|r_-|\right)+\mathbb{I}_2\,m_0|r_-|^2\right]=-(\Gamma^\mu \partial_\mu+\mathbb{I}_2 \,m_0)\frac{e^{-m_0 |r_-|}}{4\pi |r_-|},
\end{equation}
which is indeed the propagator for a massive Dirac fermion with a uniform mass $-m_0$ everywhere.

For $y=y'=0$, the boundary propagator at finite $m_0$ is
\begin{align}
\label{eq: finite m0 bdry propagator}
\begin{split}
\langle \psi(r)\, \bar{\psi}(0)\rangle|_{y=0}&=\frac{(r_-)_a\Gamma^a}{4\pi |\rho|^3}+\frac{(r_-)_a \Gamma^a \Gamma^y}{4\pi |\rho|^2}m_0\, I_1\left(\frac{m_0\,|\rho|}{2}\right) K_1\left(\frac{m_0\,|\rho|}{2}\right)\\
&\hspace{5mm}+\frac{e^{-m_0 |\rho|}(1+m_0\,|\rho|)-1}{4\pi \,m_0\,|\rho|^3}\mathbb{I}_2.
\end{split}
\end{align}
For finite $m_0$, the fermion is not purely chiral at the interface. In the $m_0\to\infty$ limit, Eq.~\eqref{eq: finite m0 bdry propagator} becomes scale-invariant and obeys a chiral Dirac matrix condition, reproducing Eq.~\eqref{eq: boundary propagator}.

If we calculate the polarization tensor $\Pi^{\mu\nu}(r,r')$ for $y\geq 0$, $y'\geq 0$ at finite $m_0$ and then take $m_0\to\infty$, we obtain a contact term,
\begin{equation}
    \Pi^{\mu\nu}(r,r')=-\,\frac{i}{4\pi}\,\varepsilon^{\mu\nu\lambda}\,\partial_\lambda \,\delta^{(3)}(r-r'),
\end{equation}
which leads to a Chern-Simons response for $y>0$ of
\begin{equation}
\label{eq: massive response}
    S_\mathrm{eff}[A_\mu]=-\,\frac{i}{8\pi}\int_{y\, >\, 0} A\, dA.
\end{equation}
Thus, if the $y>0$ side of the wall is to be regarded as a trivial insulator with Chern number $C=0$, then we must choose a massive fermion doubler that contributes the response,
\begin{equation}
    S_\mathrm{PV}[A_\mu]=\frac{i}{8\pi}\int_{\mathbb{R}^3} A\, dA,
\end{equation}
which cancels Eq.~\eqref{eq: massive response}. This point will be important in Section~\ref{sec: anomaly inflow}.

\section{Derivation of propagator near criticality}

\label{sec: near criticality propagator calc}

Here, we provide the technical details for the derivation of the fermion propagator presented in Section~\ref{sec: chern insulator near criticality}. In parallel with Sections~\ref{sec: step function mass} and \ref{sec: method of images}, we must solve the eigenvalue equation, Eq.~\eqref{eq: dirac eigenvalue eq}, but for the mass profile $M(y)$ given in Eq.~\eqref{eq: mass sign change}.

We proceed as in Section~\ref{sec: step function mass}. Again, we may write the eigenmodes in the form of Eq.~\eqref{eq: (2+1)d eigenmodes} because of translation invariance in the directions parallel to the interface of the mass domain wall, which reduces the eigenvalue equation to the form in Eq.~\eqref{eq: matrix eigenvalue eq}.

One important distinction from Section~\ref{sec: step function mass} is that for a sign change in the mass as in Eq.~\eqref{eq: mass sign change}, there is a bound state corresponding to the localized chiral edge mode in the Chern insulator phase,
\begin{equation}
    \phi_\mathrm{bound}(y)=\sqrt{\frac{2 \hspace{.25mm} m\hspace{.25mm} m_0}{m+m_0}}\begin{pmatrix}
        0\\ e^{m y}\, \Theta(-y)+ e^{-m_0 y}\, \Theta(y)
    \end{pmatrix},
\end{equation}
which has eigenvalue $\lambda=p_t+p_x$. Based on Eq.~\eqref{eq: propagator spectral}, this bound state leads to a contribution to the fermion two-point function in the $m_0\to\infty$ limit given by
\begin{equation}
    S_\mathrm{bound}(t,x;y,y')=2\hspace{.25mm}i\hspace{.25mm}m\int \frac{d^2 p}{(2\pi)^2} \frac{ e^{m(y+y')} e^{-i(p_t t+p_x x)}}{p_t+p_x-i\hspace{.25mm}\epsilon \,\mathrm{sgn}(p_x)}\begin{pmatrix}
        0&0\\
        1&0
    \end{pmatrix},
\end{equation}
which is chiral and exponentially localized at the interface of the mass domain wall.

For scattering states, the calculation proceeds much in the same way as in Section~\ref{sec: step function mass}. The eigenvalues may be parametrized as
\begin{equation}
\lambda=p_t+\sigma \sqrt{p_x^2+q^2+m^2},
\end{equation}
where $q\geq 0$ and $\sigma=\pm 1$. Since we will ultimately take the $m_0\to\infty$ limit, we consider solutions such that $0<q< \sqrt{m_0^2-m^2}$. For $y<0$, the solution is
\begin{equation}
\label{eq: finite m scattering y<0}
\phi_{\sigma,q}(y)=\frac{1}{2\sqrt{\pi\, \omega_q(\omega_q+\sigma\, p_x )}}\begin{pmatrix}
		(i\hspace{.25mm}q-m)\,e^{iqy}-R_q\,(i\hspace{.25mm}q+m)\,e^{-iq y}\\(p_x+\sigma\, \omega_q)(e^{iqy}+R_q\, e^{-iq y})
	\end{pmatrix},
\end{equation}
where in this appendix we define
\begin{equation}
    \omega_q=\sqrt{p_x^2+q^2+m^2}, \qquad R_q=\frac{i\hspace{.25mm}q-\beta_q}{i\hspace{.25mm}q+\beta_q}, \qquad \beta_q=m+m_0-\kappa_q, \qquad \kappa_q=\sqrt{m_0^2-m^2-q^2}.
\end{equation}
For $y>0$, the solution is
\begin{equation}
\label{eq: finite m scattering y>0}
	\phi_{\sigma,q}(y)=\frac{1}{2\sqrt{\pi\, \omega_q(\omega_q+\sigma\, p_x )}}\,\frac{2\hspace{.25mm}i\hspace{.25mm}q}{\beta_q + i\hspace{.25mm}q} \begin{pmatrix}
		m_0-\kappa_q\\ p_x+\sigma\, \omega_q
	\end{pmatrix}e^{-\kappa_q y}.
\end{equation}
As in Eq.~\eqref{eq: normalization}, the eigenmodes are normalized so that
\begin{equation}
\label{eq: near criticality normalization}
	\int_{-\infty}^{\infty}dy \, \phi^\dagger_{\sigma, q}(y)\,\phi_{\sigma',q'}(y)=\delta_{\sigma \sigma'}\,\delta(q-q').
\end{equation}
As a simple consistency check, for $m=0$ we observe that Eq.~\eqref{eq: finite m scattering y<0} reduces to Eq.~\eqref{eq: scattering y<0}, and Eq.~\eqref{eq: finite m scattering y>0} reduces to Eq.~\eqref{eq: scattering y>0}.

Next, we take the $m_0\to\infty$ limit. The scattering eigenmode solutions become
\begin{equation}
\phi_{\sigma, q}(y)=\frac{\Theta(-y)}{\sqrt{\pi \,\omega_q(\omega_q+\sigma\, p_x )}}\begin{pmatrix}
-(q+i\hspace{.25mm}m)\sin(q y)\\(p_x+\sigma\, \omega_q)\frac{m \sin (q y)+q \cos (q y)}{q-i \hspace{.25mm}m}
\end{pmatrix}.
\end{equation}
Using Eq.~\eqref{eq: propagator spectral}, we find that the contribution to the fermion two-point function from the scattering states is
\begin{align}
\begin{split}
&S_\mathrm{scatter}(t,x;y,y')\\
&\hspace{10mm}=\frac{2\hspace{.25mm}i}{\pi}\int_{0}^{\infty}\dd{q}\int \frac{d^2 p	}{(2\pi)^2}\frac{e^{-i(p_t t+p_x x)}}{\omega_q^2-p_t^2-i\hspace{.25mm}\epsilon}\begin{pmatrix}
-\sin (q y) f(q,y')&\hspace{5mm}-(p_t+p_x) \sin (q y) \sin (q y')\\
\frac{ (p_x-p_t)f(q,y)f(q,y')}{  m^2+q^2}&- f(q,y)\sin (q y')
\end{pmatrix},
\end{split}
\end{align}
where $f(q,y)=m \sin(q y)+q \cos(q y)$. Upon Wick rotating and calculating the integrals, the fermion propagator in Euclidean signature is given by Eq.~\eqref{eq: finite m propagator}.

\section{Anomaly coefficient of delocalized chiral current}

\label{sec: delocalized anomaly}

In this appendix, we show how to determine the anomaly coefficient of the delocalized chiral current from
\begin{equation}
    \Pi^{y\nu}_{\mathrm{odd}}(r,r')\equiv \frac{1}{|r_-|^4}\,i\hspace{.25mm}\varepsilon^{yab} \, X_a \,{I_{b}}^\nu(r_-)\,H(v)=-\,\frac{2\hspace{.25mm}i\hspace{.25mm}y}{|r_-|^5\, |r_+|}\,\varepsilon^{y\nu a}\,(\rho_a-\rho'_a)\, H(v),
\end{equation}
where $(r_\pm)_\mu=(\tau-\tau',x-x',y\pm y')$, $\rho_a=(\tau,x)$, and $a,b\in\lbrace \tau,x\rbrace$. As $y\to 0^-$, $\Pi^{y\nu}_{\mathrm{odd}}$ seems to vanish, but we will show more carefully that this limit actually yields a contact term with an anomaly coefficient. To determine the contact term carefully, we place the boundary at $y=y'=-\epsilon$ and then take the $\epsilon\to 0^+$ limit.

We Fourier transform in the directions parallel to the boundary, giving
\begin{align}
\begin{split}
    \widetilde{\Pi}^{y\nu}_\mathrm{odd}(p_\tau, p_x;y,y')&=\int d^2 \rho \, e^{-i p\cdot \rho}\,\Pi^{y\nu}_{\mathrm{odd}}(r,r')|_{\rho'=0}\\
    &=-\,\frac{\pi}{4}\, \frac{\varepsilon^{y\nu a}\,p_a}{|p|}\,\frac{1}{y\,(y')^2}\int_{v_0}^1 dv\, \frac{1-v^2}{v^4}\,\rho(v;y,y')\,J_1(\rho(v;y,y')|p|)\,H(v),
\end{split}
\end{align}
where $J_1$ is a Bessel function and
\begin{equation}
    |p|=\sqrt{p_\tau^2+p_x^2}, \qquad v_0=\left| \frac{y-y'}{y+y'} \right|,\qquad \rho(v;y,y')=\sqrt{\frac{v^2(y+y')^2-(y-y')^2}{1-v^2}}.
\end{equation}
We then integrate $\widetilde{\Pi}^{y\nu}_\mathrm{odd}$ against a test function $\Phi(y')$,
\begin{align}
\begin{split}
    \int_{-\infty}^{-\,\epsilon} dy'&\, \Phi(y')\,\widetilde{\Pi}^{y\nu}_\mathrm{odd}(p_\tau,p_x;y=-\,\epsilon,y')\\
    &=\frac{\pi}{4}\, \frac{\varepsilon^{y\nu a}\,p_a}{|p|}\int_1^\infty du\, \frac{\Phi(-\,\epsilon\, u)}{\epsilon\, u^2}\int_{\frac{u-1}{u+1}}^1 dv \,\frac{1-v^2}{v^4}\,\rho(v;1,u)\,J_1(\epsilon\,\rho(v;1,u)\,|p|)\,H(v),
    \end{split}
\end{align}
where we made the change of variables $y'=-\,\epsilon\, u$. Taking the $\epsilon\to 0^+$ limit gives
\begin{align}
\begin{split}
    \int_{-\infty}^0 dy'&\, \Phi(y')\,\widetilde{\Pi}^{y\nu}_\mathrm{odd}(p_\tau,p_x;y\to 0^-,y')\\
    &=\frac{\pi}{8}\, \varepsilon^{y\nu a}\,p_a\,\Phi(0)\int_1^\infty \frac{du}{u^2} \int_{\frac{u-1}{u+1}}^1 dv \,\frac{1-v^2}{v^4}\,[\rho(v;1,u)]^2\,H(v).
    \end{split}
\end{align}
Switching the order of integration for $u$ and $v$, we obtain
\begin{align}
\begin{split}
    \int_{-\infty}^0 dy'&\, \Phi(y')\,\widetilde{\Pi}^{y\nu}_\mathrm{odd}(p_\tau,p_x;y\to 0^-,y')\\
    &=\frac{\pi}{8}\, \varepsilon^{y\nu a}\,p_a\,\Phi(0)\int_0^1 dv \int_{1}^{\frac{1+v}{1-v}}du \,\frac{v^2(u+1)^2-(u-1)^2}{u^2\,v^4}\,\,H(v)\\
    &=\frac{\pi}{2}\, \varepsilon^{y\nu a}\,p_a\,\Phi(0)\int_0^1 dv \,\frac{(1+v^2)\,\operatorname{arctanh}(v)-v}{v^4}\,H(v) .
    \end{split}
\end{align}
In position space, we then have
\begin{equation}
    \Pi^{y\nu}_{\mathrm{odd}}(r,r')|_{y\to 0^-}=-\,\frac{i\hspace{.25mm}k_H}{4\pi}\, \varepsilon^{y\nu a}\,\partial_a \,\delta^{(2)}(\rho-\rho')\,\delta_-(y').
\end{equation}
The anomaly coefficient $k_H$ is
\begin{align}
\label{eq: anomaly coefficient}
\begin{split}
    k_H&=2\pi^2\int_0^1 \frac{dv}{v^4} \left[(1+v^2)\,\operatorname{arctanh}(v)-v\right]H(v)\\
    &=\pi^2\int_0^\infty \frac{d\xi}{\xi^2} \left[\frac{1+2\,\xi}{\sqrt{\xi(1+\xi)}}\,\operatorname{arcsinh}\sqrt{\xi}-1\right]H\left(\sqrt{\frac{\xi}{1+\xi}}\right)\\
    &=2\pi^2\int_0^\infty d\theta\, \frac{\theta(1+\coth^2\theta)-\coth\theta}{\sinh^2\theta}\,H(\tanh\theta),
    \end{split}
\end{align}
where $\xi=v^2/(1-v^2)$ and $\theta=\operatorname{arctanh}(v)$. For $H(v)=\pm\, v^2/(4\pi^2)$, we find
\begin{equation}
    k_H=\pm\,\frac{1}{2},
\end{equation}
as discussed in Section~\ref{sec: current correlation functions}.

If we had instead placed the bulk CFT in the upper half-space with $y\geq 0$ and $y'\geq 0$, we would have found
\begin{equation}
    \Pi^{y\nu}_{\mathrm{odd}}(r,r')|_{y\to 0^+}=\frac{i\hspace{.25mm}k_H}{4\pi} \,\varepsilon^{y\nu a}\,\partial_a\, \delta^{(2)}(\rho-\rho')\,\delta_+(y'),
\end{equation}
where $k_H$ is the same as given in Eq.~\eqref{eq: anomaly coefficient} and $\delta_+(y)$ is defined so that
\begin{equation}
    \int_0^\infty dy\, \Phi(y)\, \delta_+(y)=\Phi(0)
\end{equation}
for some function $\Phi(y)$. Following the analysis in Section~\ref{sec: anomaly inflow}, we then find
\begin{equation}
    \partial_\mu[\Theta(y)\,\Pi^{\mu\nu}_{\mathrm{odd}}(r,r')]=\frac{i\hspace{.25mm}k_H}{4\pi} \,\varepsilon^{y\nu a}\,\partial_a \,\delta^{(2)}(\rho-\rho')\,\delta(y)\,\delta_+(y'),
\end{equation}
so the sign of the anomaly coefficient is the same regardless of whether the gapless bulk is in the $y>0$ region or the $y<0$ region.

\section{Fermions in curved spacetime}

\label{sec: gravity review}

In this appendix, we summarize our conventions for curved Euclidean spacetime and review how a Dirac fermion couples to a background metric $g_{\mu\nu}$. This formalism provides a convenient way to study energy transport and thermal response~\cite{Luttinger1964,Volovik1990,Read2000,Ryu2012}.

We use Greek indices $\mu,\nu,\ldots$ for spacetime coordinates and capital Latin indices $A,B,\ldots$ for local orthonormal frame indices. The spacetime metric may be expressed in terms of local frame fields ${e_\mu}^A$ as
\begin{equation}
	g_{\mu\nu}={e_\mu}^A \, {e_\nu}^B \, \delta_{AB}, \qquad {e_\mu}^A\,{e^\mu}_B={\delta^A}_B.
\end{equation}
The Christoffel symbols are constructed from the metric as
\begin{equation}
	{\Gamma^\mu}_{\nu\lambda}=\frac{1}{2}\,g^{\mu \sigma}\left(\partial_\nu g_{\lambda\sigma}+\partial_\lambda g_{\nu \sigma}-\partial_\sigma g_{\nu\lambda}\right).
\end{equation}
The spin connection is
\begin{equation}
	{{\omega_\mu}^A}_B={e_\nu}^A\left(\partial_\mu {e^{\nu }}_B+{\Gamma^\nu}_{\mu \lambda} \,{e^\lambda }_B \right),
\end{equation}
which can be used to define a one-form,
\begin{equation}
	{\omega^A}_B={{\omega_\mu}^A}_B\, dx^\mu.
\end{equation}
This spin connection one-form may be used to define a gravitational Chern-Simons term with level $k_g$,
\begin{equation}
	i\hspace{.25mm}k_g\, \mathrm{CS}_g=\frac{i\hspace{.25mm}k_g}{192 \pi}\int \left({\omega^A}_B\wedge d{\omega^B}_A+\frac{2}{3}\,{\omega^A}_B\wedge {\omega^B}_C\wedge {\omega^C}_A\right).
\end{equation}
The corresponding chiral central charge is $c_-=k_g/2$, and the associated thermal Hall conductivity is
\begin{equation}
	\kappa_{xy}=c_-\,\frac{\pi k_B^2}{6\hbar}\,T,
\end{equation}
where $T$ is the temperature. This gravitational term is globally well-defined for $k_g\in\mathbb{Z}$. In the Chern insulator phase with Chern number $C$, the level is $k_g=2\,C$, and the chiral central charge is $c_-=C$. For the Chern insulator transition between the $C=1$ and $C=0$ states, since the effective Chern number is $C=1/2$, the gravitational response has level $k_g=1$.

We next expand the gravitational response about flat spacetime. We work in linearized gravity, taking a metric,
\begin{equation}
	g_{\mu\nu}=\delta_{\mu\nu}+h_{\mu\nu},
\end{equation}
where $|h_{\mu\nu}|\ll 1$. We take the local frame fields to be
\begin{equation}
	{e_\mu}^A={\delta_\mu}^A+\frac{1}{2}\, {h_\mu}^A,\qquad {e^\mu}_A={\delta^\mu}_A-\frac{1}{2}\, {h^\mu}_A.
\end{equation}
The Christoffel symbols to linear order in $h_{\mu\nu}$ are then
\begin{equation}
{\Gamma^\mu}_{\nu\lambda}=\frac{1}{2}\left(\partial_\nu {h_{\lambda}}^\mu+\partial_\lambda {h_{\nu }}^\mu-\partial^\mu h_{\nu\lambda}\right).
\end{equation}
The spin connection in linearized gravity is
\begin{equation}
	{{\omega_\mu}^A}_B=\frac{1}{2}\left(\partial_B \,{h_\mu}^A-\partial^A \, h_{\mu B}\right).
\end{equation}
After an integration by parts, the gravitational Chern-Simons term to quadratic order in $h_{\mu\nu}$ is
\begin{equation}
\label{eq: appendix grav CS}
i\hspace{.25mm}k_g\, \mathrm{CS}_g=\frac{i\hspace{.25mm}k_g}{384 \pi}\int d^3 r \, \varepsilon^{\mu\nu\lambda}\, h_{\mu\sigma}\, \partial_\nu \left(\partial^2 {h_\lambda}^\sigma-\partial^\sigma \partial_\rho {h_\lambda}^\rho\right).
\end{equation}
Eq.~\eqref{eq: appendix grav CS} is the form of the gravitational Chern-Simons term used in Section~\ref{sec: gravitational anomaly}.

Next, we review how to place fermionic fields in curved spacetime. The Dirac matrices in flat spacetime $\gamma^A$ satisfy\footnote{In this appendix only, we use the notation $\gamma^A$ instead of $\Gamma^A$ for Dirac matrices in Euclidean signature to avoid confusion with the Christoffel symbols.}
\begin{equation}
	\left\lbrace \gamma_A,\gamma_B\right\rbrace =2\, \delta_{AB}.
\end{equation}
The generators of $\mathrm{Spin}(3)\cong SU(2)$, the double cover of $SO(3)$, are
\begin{equation}
	\sigma_{AB}=\frac{i}{4}\,[\gamma_A,\gamma_B],
\end{equation}
and obey the algebra,
\begin{equation}
[\sigma_{AB},\sigma_{CD}]=i\left(\delta_{AD}\,\sigma_{BC}+\delta_{BC}\,\sigma_{AD}-\delta_{AC}\,\sigma_{BD}-\delta_{BD}\,\sigma_{AC}\right).
\end{equation}
To couple a Dirac fermion $\psi$ to a background metric $g_{\mu\nu}$, we introduce the covariant derivative,
\begin{equation}
	\nabla_\mu\psi=\left(\partial_\mu -\frac{i}{2}\, {\omega_\mu}^{AB}\,\sigma_{AB}\right)\psi.
\end{equation}
Additionally, we define
\begin{equation}
	\slashed{\nabla}={e^\mu}_A\,\gamma^A \,\nabla_\mu.
\end{equation}
The action of the fermion coupled to the background metric is
\begin{equation}
	S=\int  d^3r \,\sqrt{g}\, \bar{\psi} \left(i\slashed{\nabla}+m\right)\psi,
\end{equation}
where $g$ is the determinant of $g_{\mu\nu}$ and $m$ is the fermion mass.

Integrating out the massive fermion and expanding to quadratic order in $h_{\mu\nu}$ gives a gravitational response of
\begin{align}
    S_{\mathrm{resp}}[h_{\mu\nu}]=-\, \frac{1}{8}\int d^3 r\int d^3 r'\, h_{\mu\nu}(r)\, \langle T^{\mu\nu}(r)\, T^{\lambda\sigma}(r')\rangle \, h_{\lambda\sigma}(r').
\end{align}
The energy-momentum tensor in flat spacetime is given by Eq.~\eqref{eq: energy-momentum tensor}. For a Dirac fermion of uniform mass, in the $|m|\to\infty$ limit, we obtain a universal parity-odd term~\cite{Alvarez-Gaume1985,Goni1986,Vuorio1986},
\begin{equation}
   \langle T_{\mu\nu}(r)\, T_{\lambda\sigma}(r')\rangle=-\,\frac{i\hspace{.25mm}\operatorname{sgn}(m)}{192\pi}\left\lbrace \left[\varepsilon_{\mu\lambda\rho} \,\partial^\rho\,(\partial_\nu \,\partial_\sigma-\delta_{\nu\sigma}\,\partial^2 )+(\mu \leftrightarrow\nu)\right]+(\lambda\leftrightarrow\sigma)\right\rbrace \delta^{(3)}(r-r').
\end{equation}
This contact term leads to a response,
\begin{equation}
\label{eq: appendix gCS response}
    S_{\mathrm{gCS}}[h_{\mu\nu}]=\frac{i\hspace{.25mm}\operatorname{sgn}(m)}{384 \pi}\int d^3 r \, \varepsilon^{\mu\nu\lambda}\, h_{\mu\sigma}\, \partial_\nu \left(\partial^2 {h_\lambda}^\sigma-\partial^\sigma \partial_\rho {h_\lambda}^\rho\right),
\end{equation}
which is a gravitational Chern-Simons term with level $k_g=\operatorname{sgn}(m)$. In Section~\ref{sec: gravitational anomaly}, the heavy fermion doubler then has a contribution of Eq.~\eqref{eq: appendix gCS response} with $m>0$.

\section{(3+1)d mass domain wall}

\label{sec: (3+1)d mass domain wall}

In this appendix, we derive the correlation function of a Dirac fermion in (3+1)d with a mass domain wall configuration as given in Eq.~\eqref{eq: (3+1)d mass domain wall}, as was similarly done in Ref.~\cite{Mulligan2013}. In analogy with Section~\ref{sec: step function mass}, we must solve the eigenvalue equation,
\begin{equation}
	\gamma^0\,[i\hspace{.25mm} \gamma^\mu\, \partial_\mu-M(z)]\,\varphi_\lambda(t,x,y,z)=\lambda \, \varphi_\lambda(t,x,y,z),
\end{equation}
where $M(z)=-\,m_0 \,\Theta(z)+m\,\Theta(-z)$ and $0<m<m_0$. Using translation invariance in the directions parallel to the boundary, we take
\begin{equation}
	\varphi_\lambda(t,x,y,z)=\frac{1}{(2\pi)^{3/2}}e^{-i (p_t t+p_x x+p_y y)}\phi_\lambda(z).
\end{equation}
Our basis for the 4 by 4 Dirac matrices in Lorentzian signature consists of
\begin{align}
	\gamma^0=\begin{pmatrix}
		0&-i\hspace{.25mm} \sigma^1\\i\hspace{.25mm} \sigma^1&0
	\end{pmatrix}, \qquad \gamma^1=\begin{pmatrix}
	0&\sigma^2\\-\,\sigma^2&0
	\end{pmatrix}, \qquad \gamma^2=\begin{pmatrix}
	0&\sigma^3\\-\,\sigma^3&0
	\end{pmatrix}, \qquad \gamma^3=\begin{pmatrix}
	i\hspace{.25mm} \mathbb{I}_2&0\\0&-\,i\hspace{.25mm} \mathbb{I}_2
	\end{pmatrix}.
\end{align}
In this basis, the eigenvalue equation then reduces to
\begin{equation}
\label{eq: (3+1)d eigenvalue eq}
\begin{pmatrix}
	-\,\sigma^3 \,p_x+\sigma^2\, p_y &-\,i \hspace{.25mm} \sigma^1\,[d/dz-M(z)] \\
-\,i\hspace{.25mm} \sigma^1\,[d/dz+M(z)]	& 	-\,\sigma^3 \,p_x+\sigma^2 \, p_y
\end{pmatrix}\phi_\lambda(z)=(\lambda-p_t)\,\phi_\lambda(z),
\end{equation}
which may be solved in a manner similar to the eigenvalue equation in Section~\ref{sec: near criticality propagator calc}.

We may form a normalized basis of bound state solutions to Eq.~\eqref{eq: (3+1)d eigenvalue eq} from
\begin{align}
\label{eq: (3+1)d bound state}
\begin{split}
	\phi^\mathrm{bound}_{\sigma}(z)&=\sqrt{\frac{2m_0 m}{m_0+m}}[e^{m z}\,\Theta(-z)+e^{-m_0 z}\,\Theta(z)]\begin{pmatrix}
	0\\	\chi_\sigma
	\end{pmatrix}, \\ \chi_\sigma &=\frac{1}{\sqrt{2p_\parallel(p_\parallel+\sigma \, p_x)}}\begin{pmatrix}
	-\,i\hspace{.25mm} p_y\\ p_x+\sigma \,p_\parallel
	\end{pmatrix},
    \end{split}
\end{align}
which has a corresponding eigenvalue of $\lambda=p_t+\sigma \, p_\parallel$, where $\sigma=\pm 1$ and $p_\parallel=\sqrt{p_x^2+p_y^2}$. The two-component spinor $\chi_\sigma$ satisfies
\begin{equation}
    h(p_x,p_y)\, \chi_\sigma= \sigma\, p_\parallel \, \chi_\sigma, \qquad h(p_x,p_y)=-\,\sigma^3\, p_x+\sigma^2 \, p_y.
\end{equation}
Here, $h(p_x,p_y)$ is the Hamiltonian for a massless (2+1)d Dirac fermion in momentum space.

Next, we consider scattering state solutions to Eq.~\eqref{eq: (3+1)d eigenvalue eq}, which have eigenvalues that may be parametrized as
\begin{equation}
    \lambda=p_t+\sigma \, \omega_q,
\end{equation}
where $\omega_q=\sqrt{p_\parallel^2+q^2+m^2}$ and $q>0$. Since we eventually take $m_0\to\infty$, we only consider $0<q<\sqrt{m_0^2-m^2}$. For $z<0$, the solution is
\begin{equation}
	\phi_{\sigma,s, q}(z)=\mathcal{N}_{\sigma,s}\begin{pmatrix}
		[q \cos(qz)-(m+m_0+\kappa_q)\sin(qz)]\,\chi_s\\
		i \frac{m_0+\kappa_q}{\sigma\,\omega_q+s\, p_\parallel}[q\cos(qz)+(m+m_0-\kappa_q)\sin(qz)]\,\sigma^1\,\chi_s
	\end{pmatrix},
\end{equation}
where $\kappa_q=\sqrt{m_0^2-m^2-q^2}$ and $s=\pm 1$ is a polarization that may be chosen independently of $\sigma$. The solution for $z>0$ is
\begin{equation}
	\phi_{\sigma,s,q}(z)=\mathcal{N}_{\sigma,s} \, q \, e^{-\kappa_q z} \begin{pmatrix}
		\chi_s \\  i \frac{m_0+\kappa_q}{\sigma\, \omega_q+s\,p_\parallel}\,\sigma^1 \,\chi_s
	\end{pmatrix}.
\end{equation}
The normalization factor,
\begin{equation}
\mathcal{N}_{\sigma,s}=\sqrt{\frac{\omega_q+\sigma \,s \,p_\parallel}{2\pi \,\omega_q\,(m+m_0)\,(m_0+\kappa_q)}},
\end{equation}
is chosen so that
\begin{equation}
    \int_{-\infty}^\infty dz\, \phi_{\sigma,s,q}^\dagger(z)\,\phi_{\sigma',s',q'}(z)=\delta_{\sigma \sigma'}\,\delta_{s s'}\, \delta(q-q'),
\end{equation}
in analogy with Eq.~\eqref{eq: near criticality normalization}.

In the $m_0\to\infty$ limit, the bound state solution, Eq.~\eqref{eq: (3+1)d bound state}, reduces to
\begin{equation}
	\phi^\mathrm{bound}_{\sigma}(z)=\sqrt{2m}\, e^{m z}\,\Theta(-z)\begin{pmatrix}
	0\\	\chi_\sigma
\end{pmatrix},
\end{equation}
which leads to a contribution to the fermion two-point function (in Euclidean signature) of
\begin{equation}
	S_\mathrm{bound}^\mathrm{4d}(r_\parallel;z,z')=\frac{m \,e^{m(z+z')}\,\Gamma^a \,(r_\parallel)_a}{2\pi |r_\parallel|^3}\frac{\mathbb{I}_4+\Gamma^z}{2}=\frac{m\, e^{m(z+z')}}{2\pi |r_\parallel|^3}\begin{pmatrix}
		0&0\\ i\hspace{.25mm}\sigma^a \,(r_\parallel)_a&0
	\end{pmatrix},
\end{equation}
where $a\in \lbrace \tau ,x,y\rbrace$, $(r_\parallel)_a=(\tau,x,y)$, and $|r_\parallel|=\sqrt{\tau^2+x^2+y^2}$. This contribution describes the (2+1)d Dirac cone localized at the (2+1)d surface of the (3+1)d topological insulator.

In the $m_0\to\infty$ limit, the scattering state solutions become
\begin{equation}
	\phi_{\sigma,s, q}(z)=\frac{1}{\sqrt{\pi \,\omega_q}}\begin{pmatrix}
		-\sqrt{\omega_q+\sigma \,s\, p_\parallel} \sin(q z) \,\chi_s\\ \sigma\,\frac{ i}{ \sqrt{\omega_q +\sigma \,s\, p_\parallel}}[q\cos(qz)+m\sin(qz)]\,\sigma^1 \,\chi_s
	\end{pmatrix}.
\end{equation}
The contribution to the fermion two-point function (in Euclidean signature) from the scattering states is
\begin{equation}
	S_\mathrm{scatter}^\mathrm{4d}(r_\parallel;z,z')=\int \frac{d^3 p_\parallel}{(2\pi)^3}\frac{2}{\pi}\int_{0}^{\infty}dq\,\frac{e^{-ip_a (r_\parallel)^a}}{\omega_q^2+p_\tau^2}\begin{pmatrix}
		\sin(q z) f(q,z')\,\mathbb{I}_2& \hspace{4mm}\sin(q z)\sin(q z')\,\sigma^a\, p_a\\
		-\,\frac{ f(q,z)\,f(q,z')}{q^2+m^2}\,\sigma^a \,p_a&f(q,z)\sin(q z')\,\mathbb{I}_2
	\end{pmatrix},
\end{equation}
where $f(q,z)=q \cos(q z)+m\sin(q z)$ and $\sigma^a \,p_a=\sigma^1\, p_\tau +\sigma^2 \, p_x+\sigma^3 \,p_y$.

Performing the integrals, we find that the fermion propagator for finite $m$ (but $m_0\to\infty$) is
\begin{align}
\label{eq: 4d massive propagator}
	\begin{split}
	S_E^\mathrm{4d}(r,r')&=S_m^\mathrm{4d}(r_-)+S_m^\mathrm{4d}(r_+)\,\Gamma^z-\frac{m^2 \,K_1(m |r_+|)}{4\pi^2 \,|r_+|}\,(\mathbb{I}_4+\Gamma^z)+\frac{m\, e^{m(z+z')}\,\Gamma^a \, (r_-)_a}{2\pi |r_\parallel|^3}\frac{\mathbb{I}_4+\Gamma^z}{2}\\
	&\hspace{5mm}+\frac{\Gamma^a (r_-)_a}{|r_\parallel|}\frac{\mathbb{I}_4+\Gamma^z}{2}\partial_{|r_\parallel|} (\partial_z+m)\,\mathcal{K}(m,|r_\parallel|,z+z'),
	\end{split}
\end{align}
where $(r_\pm)_\mu=(\tau-\tau',x-x',y-y',z\pm z')$, the propagator for a Dirac fermion of uniform mass $m$ everywhere in (3+1)d Euclidean spacetime is
\begin{equation}
	S_m^{\mathrm{4d}}(r)=(-\,\Gamma^\mu \,\partial_\mu+m \,\mathbb{I}_4)\frac{m\, K_1(m |r|)}{4\pi^2 \,|r|}=\frac{m^2[\Gamma^\mu \, r_\mu \,K_2(m |r|)+|r| \, K_1(m|r|)]}{4\pi^2 \, |r|^2},
\end{equation}
and the remaining integral is
\begin{equation}
\mathcal{K}(m,|r_\parallel|,z+z')=\frac{1}{2\pi^2 \,|r_\parallel|}\int_{0}^{\infty} du\frac{e^{-\,m \,|r_\parallel|\,\sqrt{1+u^2}}\cos[m\,(z+z')\,u]}{1+u^2}.
\end{equation}
In the $m\to 0$ limit, the propagator in Eq.~\eqref{eq: 4d massive propagator} reduces to Eq.~\eqref{eq: TI propagator}. 

We note that the propagator, Eq.~\eqref{eq: 4d massive propagator}, is not a simple method of images solution. As in the (2+1)d case in Section~\ref{sec: chern insulator near criticality} and Appendix~\ref{sec: near criticality propagator calc}, the naive image solution will not satisfy the boundary conditions because a massive fermion is not invariant under
\begin{equation}
    \psi(\tau,x,y,z)\to \zeta_z \, \Gamma^z\, \psi(\tau,x,y,-z).
\end{equation}
However, a (3+1)d massive Dirac fermion is invariant under
\begin{equation}
\label{eq: (3+1)d reflection}
    \psi(\tau,x,y,z)\to \zeta_z \, \Gamma^z\, \Gamma^5\, \psi(\tau,x,y,-z),
\end{equation}
and there is indeed a method of images solution even for finite $m$ if we instead imposed the boundary condition,
\begin{equation}
i\hspace{.25mm}\Gamma^z \, \Gamma^5 \psi(\tau,x,y,z=0)=\psi(\tau,x,y,z=0),
\end{equation}
since a massive Dirac fermion is invariant under this reflection symmetry, Eq.~\eqref{eq: (3+1)d reflection}.

Now that we have the fermion propagator at the TI to trivial insulator transition, as in Section~\ref{sec: current correlation functions}, we may also use it to compute correlation functions of the $U(1)$ current, $J^\mu(r)=-\, i \hspace{.25mm}\bar{\psi}(r)\, \Gamma^\mu \, \psi(r)$. At the critical point, $m=0$, the current two-point function is
\begin{equation}
\label{eq: TI transition currents}
	\langle J_\mu(r)\,J_\nu(r')\rangle =\frac{1}{|r_-|^6}\left[I_{\mu\nu}(r_-) \,\frac{1+v^6}{\pi^4}-X_\mu \, X_\nu ' \,\frac{2\hspace{.25mm}v^6}{\pi^4}\right],
\end{equation}
where $I_{\mu\nu}(r_-)$ is defined in Eq.~\eqref{eq: I tensor}, $v=|r_-|/|r_+|$, and 
\begin{gather}
	X_\mu= v\left(\frac{2\hspace{.25mm}z}{|r_-|^2}\,(r_-)_\mu-\delta_{\mu z}\right), \qquad  X'_\mu =-\,v\left(\frac{2\hspace{.25mm}z'}{|r_-|^2}\,(r_-)_\mu+\delta_{\mu z}\right).
\end{gather}
Unlike the current correlation function for the Chern insulator transition in (2+1)d (cf. Eq.~\eqref{eq: cft chern transition currents}), the current two-point function in Eq.~\eqref{eq: TI transition currents} is time-reversal invariant. 
	\bibliographystyle{SciPost_bibstyle}
	\bibliography{references.bib}
	
\end{document}